\documentclass[10pt,prd,twocolumn,
nofootinbib,
noshowkeys,
noshowpacs,
superscriptaddress,
floatfix
]{revtex4-2}
\usepackage{amsmath,amsfonts,amsthm,amssymb}
\usepackage[dvips]{graphics,graphicx}
\usepackage[usenames,dvipsnames]{color}
\definecolor{darkblue}{RGB}{0,0,196}
\definecolor{darkgreen}{RGB}{0,120,0}
\usepackage[colorlinks=true,linktocpage=true,linkcolor=darkblue,citecolor=red,urlcolor=darkblue]{hyperref}
\usepackage{cancel}
\usepackage{bbold}
\usepackage{multirow}
\usepackage{longtable}
\usepackage{color}
\usepackage[normalem]{ulem}
\usepackage{hyperref}
\usepackage{bigints}
\usepackage{xparse}
\usepackage{physics}
\usepackage{verbatim}
\usepackage{minibox}
\usepackage{comment}
\usepackage{appendix}
\usepackage{slashed}
\usepackage{marginnote}
\usepackage{graphicx}
\usepackage[nice]{nicefrac}
\usepackage{amsmath}
\usepackage{siunitx}
\usepackage{hepunits}
\usepackage{float}

\usepackage{stackengine,scalerel}
\newcommand\hstar[1]{\ThisStyle{\ensurestackMath{%
  \setbox0=\hbox{$\SavedStyle#1$}%
  \stackengine{0pt}{\copy0}{\kern.2\ht0\smash{\SavedStyle\star}}{O}{c}{F}{T}{S}}}}

\definecolor {darkgreen}{rgb}{0.2,0.7,0.2}

\begin{document}

\title{Dissipative spin hydrodynamics in Bjorken flow and thermal dilepton production}

\author{Sejal Singh}
\email{p20230472@pilani.bits-pilani.ac.in}
\affiliation{Department of Physics,  Birla Institute of Technology and Science Pilani, Pilani Campus, Pilani,  Rajasthan-333031, India}

\author{Sourav Dey}
\email{sourav.dey@niser.ac.in}
\affiliation{Department of Physics,  Birla Institute of Technology and Science Pilani, Pilani Campus, Pilani,  Rajasthan-333031, India}
\affiliation{School of Physical Sciences, National Institute of Science Education and Research, An OCC of Homi Bhabha National Institute, Jatni-752050, India}

\author{Arpan Das}
\email{arpan.das@pilani.bits-pilani.ac.in}
\affiliation{Department of Physics,  Birla Institute of Technology and Science Pilani, Pilani Campus, Pilani,  Rajasthan-333031, India}

\author{Hiranmaya Mishra}
\email{hiranmaya@niser.ac.in}
\affiliation{School of Physical Sciences, National Institute of Science Education and Research, An OCC of Homi Bhabha National Institute, Jatni-752050, India}
\affiliation{Institute of Physics, Sachivalaya Marg, Bhubaneswar-751005, India}

\author{Amaresh Jaiswal}
\email{a.jaiswal@niser.ac.in}
\affiliation{School of Physical Sciences, National Institute of Science Education and Research, An OCC of Homi Bhabha National Institute, Jatni-752050, India}

\begin{abstract}
We investigate the boost-invariant expansion of a recently developed first-order spin hydrodynamic framework in which the spin chemical potential is treated as a leading-order hydrodynamic variable. Considering a symmetric energy-momentum tensor and a separately conserved spin tensor, we derive the coupled evolution equations for the medium temperature and the independent components of the spin chemical potential in the presence of both viscous and spin-diffusive transport coefficients. For a boost-invariant system, only the magnetic-like components of the spin chemical potential survive, and their evolution is shown to depend sensitively on the spin transport coefficients. The transverse spin components decay more rapidly due to spin dissipation, while the longitudinal component survives for a longer duration. We further demonstrate that the evolution of the spin degrees of freedom modifies the temperature profile of the expanding medium. Using the resulting temperature profiles, we calculate thermal dilepton production rates from quark-antiquark annihilation. We find that the presence of spin dynamics enhances the dilepton yield relative to standard dissipative hydrodynamics, with the magnitude of the enhancement depending on the spin transport coefficients. Within the simplified hydrodynamic evolution considered here, our results qualitatively indicate that thermal dileptons can possibly provide an indirect probe of spin dynamics and spin transport in the quark-gluon plasma.
\end{abstract}
	
\date{\today \hspace{0.2truecm}}

\maketitle
\flushbottom

\section{Introduction}
Spin polarization of different hadrons has been measured in heavy-ion collision experiments~\cite{STAR:2017ckg,STAR:2018gyt,STAR:2019erd,ALICE:2019onw,Kornas:2020qzi}.
Spin polarization of hadrons can be argued to originate from spin-vorticity coupling~\cite{Liang:2004ph,Becattini:2007sr,Gao:2007bc,Huang:2011ru,Becattini:2013fla,Fang:2016vpj,Voloshin:2004ha,Betz:2007kg, Becattini:2015ska,Becattini:2022zvf}.
Non-vanishing vorticity and its effects on the dynamics of the partonic medium produced in heavy-ion collisions have gained a lot of attention. The vorticity generation in the plasma produced in heavy-ion collisions is primarily associated with the large angular momentum involved in non-central heavy-ion collisions~\cite{Liang:2004ph,Becattini:2007sr,Gao:2007bc,Huang:2011ru}.
In peripheral heavy ion collisions, the large angular momentum associated with colliding nuclei can give rise to non-vanishing vorticity in the thermalized partonic medium due to the nuclei’s inhomogeneous density profile~\cite{Liang:2004ph,Becattini:2007sr,Gao:2007bc,Huang:2011ru}. As a result of the spin-vorticity coupling, different hadrons can become polarized in the vortical medium. Due to the well-understood weak decay process, hyperons are considered to be an important probe for the measurement of spin polarization~\cite{STAR:2017ckg}. Various theoretical models have been developed to explain the spin polarization measurements of Lambda ($\Lambda$), and anti-Lambda ($\bar{\Lambda}$) hyperons, e.g., relativistic dissipative hydrodynamic models~\cite{DelZanna:2013eua,Karpenko:2013wva,Ivanov:2019ern}), parton cascade model (AMPT)~\cite{Li:2017slc}, hadronic cascade model (UrQMD)~\cite{Vitiuk:2019rfv}, chiral kinetic theory~\cite{Sun:2017xhx}, etc. These models use the spin-thermal vorticity coupling to explain the {\it global spin polarization} of hyperons, i.e., the polarization along the direction of global angular momentum. However, spin-thermal vorticity coupling alone does not provide a satisfactory explanation for the {\it local spin polarization} measurement, i.e., the longitudinal (along beam direction) polarization as a function of the azimuthal angle in the transverse plane (transverse to the beam direction)~\cite{Becattini:2020ngo}. Apart from the thermal vorticity, thermal shear can also play an important role in the polarization of hyperons
~\cite{Becattini:2021suc,Liu:2021uhn,Becattini:2021iol,Fu:2021pok}.  

To explain the spin polarization measurements various groups have developed a novel theoretical approach, known as the spin hydrodynamic framework. This approach generalizes the standard hydrodynamic framework (for spin-less fluid), to incorporate the dynamical evolution of spin~\cite{Florkowski:2018fap, Florkowski:2018ahw, Florkowski:2017dyn, Florkowski:2017ruc, Florkowski:2019qdp, Florkowski:2019voj, Bhadury:2021oat}. Contrary to the standard hydrodynamic approach, where one only considers the conservation of the total energy-momentum tensor ($\partial_{\mu}T^{\mu\nu}=0$), in spin hydrodynamic frameworks one also considers the conservation of the angular momentum tensor ($\partial_{\lambda}J^{\lambda\mu\nu}=0$). The dynamical evolution of the spin degree of freedom is encoded in the conservation of the total angular momentum tensor. Different methods, e.g., entropy current analysis~\cite{Hattori:2019lfp,Fukushima:2020ucl,Li:2020eon,She:2021lhe}, the effective Lagrangian approach~\cite{Montenegro:2017lvf,Montenegro:2017rbu}, the kinetic-theory approach~\cite{Florkowski:2017ruc,Florkowski:2018myy,Bhadury:2020puc,Shi:2020qrx,Weickgenannt:2022zxs,Weickgenannt:2019dks,Weickgenannt:2020aaf}, etc. have been used to develop spin hydrodynamic frameworks. Recently, in Refs.~\cite{Singh:2024cub,Sapna:2025yss}, numerical simulations of spin hydrodynamic frameworks have also been reported.  

Using the entropy current analysis method, in Ref. ~\cite{Dey:2024cwo}, some of us have developed a first-order spin hydrodynamic theory. Such a theory can be considered as the Navier-Stokes limit of the higher-order spin hydrodynamic framework. Similar frameworks have also been discussed in Refs.~\cite{Hattori:2019lfp, Fukushima:2020ucl,Wang:2021ngp,Daher:2022xon,Sarwar:2022yzs,Daher:2022wzf}. One of the distinguishing features of our framework as compared to the framework discussed in Refs.~\cite{Hattori:2019lfp, Fukushima:2020ucl,Wang:2021ngp,Daher:2022xon,Sarwar:2022yzs,Daher:2022wzf} is the hydrodynamic treatment of spin chemical potential ($\omega^{\alpha\beta}$). In the spin hydrodynamic framework, the evolution of the spin chemical potential encodes the dynamics of the spin degree of freedom~\cite{Florkowski:2018fap,Florkowski:2018ahw,Florkowski:2017dyn}. In such a framework, the spin chemical potential is also treated as a hydrodynamic variable, along with temperature, chemical potential, etc. Hence, the hydrodynamic ordering of the spin chemical potential is crucial. In literature, one often considers that spin chemical potential is a $\mathcal{O}(\partial)$ term in the hydrodynamic gradient expansion~\cite{Hattori:2019lfp}. This is essential when one deals with the asymmetric energy-momentum tensor~\cite{Hattori:2019lfp}. But for a symmetric energy-momentum tensor, for theoretical consistency, one can conclude that spin chemical potential is a leading order term, i.e., $\mathcal{O}(1)$ term in the hydrodynamic gradient expansion~\cite{She:2021lhe,Dey:2024cwo}. Such different hydrodynamic ordering of spin chemical potential gives rise to qualitatively different spin hydrodynamic frameworks, e.g., in the framework discussed in Ref.~\cite{Hattori:2019lfp} where one considers that the spin chemical potential is $\mathcal{O}(\partial)$, the dissipative part of the spin tensor can not be fixed at the Navier-Stokes limit. However, if we consider that spin chemical potential is $\mathcal{O}(1)$, then within the first-order theory, constitutive relations of the dissipative part of the spin tensor can be obtained. The novelty of the framework developed in Ref.~\cite{Dey:2024cwo} is the spin transport coefficients and the associated Green-Kubo relations. 

In this work, we consider the spin hydrodynamic framework developed in Ref.~\cite{Dey:2024cwo}, and discuss the boost-invariant flow (Bjorken flow) solution of the spin hydrodynamic equations \footnote{ A fundamental origin of development of the spin hydrodynamics is local spin polarization, which is measured in non-central heavy ion collisions. The qualitative argument for the local spin polarization originates from the anisotropic flow on the transverse plane in the non-central heavy ion collision~\cite{STAR:2019erd}. On the other hand the implicit assumption for the Bjorken flow is that the flow is only along the longitudinal direction and there is no radial flow. In fact, it can be shown that for Bjorken flow, thermal vorticity vanishes identically. Hence, Bjorken flow can not explain the local spin polarization data using the theoretical model that incorporates the spin-thermal vorticity coupling~\cite{Becattini:2020ngo}. To explain the spin polarization measurement, one may possibly need a 3+1 D spin hydrodynamic framework with a proper initial condition~\cite{Sapna:2025yss}. In the current paper, we are not addressing the local spin polarization measurement. Rather, we are trying to find a possible solution, namely a boost invariant solution to an existing spin-hydrodynamic framework.}. In the context of relativistic hydrodynamics Bjorken flow is the standard choice, due to the boost-invariant nature of the flow at the ultra-relativistic energies. The Bjorken flow solution is also considered as a testbed for the more general 3+1 D hydrodynamic simulations. Hence, the Bjorken flow solution has been investigated for different spin hydrodynamic frameworks in literature~\cite{Florkowski:2019qdp,Wang:2021ngp,Biswas:2022bht,Drogosz:2024lkx}. In the context of spin hydrodynamics, a boost-invariant flow solution has also been used to study the attractor solutions in Ref.~\cite{Wang:2024afv}. In some references, authors have considered the Bjorken flow solution for ideal spin hydrodynamic equations, i.e., without any dissipation~\cite{Florkowski:2019qdp}. On the other hand, in Refs.~\cite{Wang:2021ngp,Biswas:2022bht} Bjorken flow solution has been discussed, but the spin diffusion, or the dissipative part of the spin dynamics, does not play any role. In the present investigation, we incorporate the dissipative parts of the energy-momentum tensor and the spin tensor. More importantly, we demonstrate  the effect of spin transport coefficients on the proper time evolution of temperature and spin chemical potential. For simplicity, we consider a baryon-free system. Subsequently, the proper time evolution of medium temperature is used to calculate the thermal dilepton production rate~\cite{Florkowski:2010zz,Vogt:2007zz,Alam:1996fd}. Thermal dileptons are an excellent probe of the medium's temperature.  Dileptons interact only electromagnetically; hence, they have a much longer mean free path than hadrons. In our calculation it is the temperature evolution of the plasma, that affects the dilepton production rate. We demonstrate that spin evolution affects the temperature evolution of a longitudinally expanding system, which also affects the dilepton production rates.

In this manuscript, we use the following notations and conventions. $u^{\mu}$ is the normalized fluid flow vector, $u^{\mu}u_{\mu}=1$. $\Delta^{\mu\nu}=g^{\mu\nu}-u^{\mu}u^{\nu}$ is the projector normal to $u^{\mu}$, i.e., $\Delta^{\mu\nu}u_{\nu}=0$.  $g_{\mu\nu}= \hbox{diag}(+1, -1, -1, -1)$ is the metric tensor. 
$A^{\langle\mu\rangle}\equiv \Delta^{\mu\nu}A_{\nu}$ is the 
projection of a four vector $A^{\mu}$ orthogonal to $u^{\mu}$. $A^{\{\alpha}B^{\beta\}}=(A^{\alpha}B^{\beta}+A^{\beta}B^{\alpha})/2$ represents the symmetric combination, and $A^{[\alpha}B^{\beta]}=(A^{\alpha}B^{\beta}-A^{\beta}B^{\alpha})/2$ represents the anti-symmetric combination. $A^{\langle\mu}B^{\nu\rangle}$ denotes traceless and symmetric projection orthogonal to fluid flow. It is defined as $A^{\langle\mu}B^{\nu\rangle}\equiv \Delta^{\mu\nu}_{\alpha\beta}A^{\alpha}B^{\beta}\equiv \frac{1}{2}\left(\Delta^{\mu}_{~\alpha}\Delta^{\nu}_{~\beta}+\Delta^{\mu}_{~\beta}\Delta^{\nu}_{~\alpha}-\frac{2}{3}\Delta^{\mu\nu}\Delta_{\alpha\beta}\right)A^{\alpha}B^{\beta}$. By definition, $u_{\mu}A^{\langle\mu}B^{\nu\rangle}=0$, $u_{\nu}A^{\langle\mu}B^{\nu\rangle}=0$, and $A^{\langle\mu}B^{\nu\rangle}=A^{\langle\nu}B^{\mu\rangle}$. Similarly, the antisymmetric projection operator orthogonal to the flow vector is defined as $A^{\langle[\mu}B^{\nu]\rangle}\equiv \Delta^{[\mu\nu]}_{[\alpha\beta]}A^{\alpha}B^{\beta}\equiv \frac{1}{2}\left(\Delta^{\mu}_{~\alpha}\Delta^{\nu}_{~\beta}-\Delta^{\mu}_{~\beta}\Delta^{\nu}_{~\alpha}\right)A^{\alpha}B^{\beta}$. Partial derivative $\partial_{\mu}$ can be decomposed along the flow direction and normal to the flow direction, $\partial_{\mu}=u_{\mu}D+\nabla_{\mu}$. Here 
$D\equiv u^{\mu}\partial_{\mu}$, that represents comoving derivative, and $\nabla_{\mu}\equiv\Delta_{\mu}^{~\alpha}\partial_{\alpha}$. $\theta\equiv \partial_{\mu}u^{\mu}=\nabla_{\mu}u^{\mu}$ is the fluid expansion rate. 
$\sigma_{\mu\nu}$ is the symmetric traceless combination of the derivative of the fluid flow. It is defined as, 
$\sigma_{\mu\nu}\equiv\frac{1}{2}(\nabla_{\mu} u_{\nu}+\nabla_{\nu} u_{\mu})-\frac{1}{3}\theta\Delta_{\mu\nu}$.
$\epsilon^{\mu\nu\alpha\beta}$ is the totally antisymmetric Levi-Civita tensor with the sign convention $\epsilon^{0123} = -\epsilon_{0123} = 1$. 

The rest of the manuscript is organized in the following way. In Sec .~\ref{SecII} we discuss the dissipative spin hydrodynamic framework. Here, we also discuss the spin-dependent equation of state. In Sec .~\ref{SecIII}, and Sec .~\ref{SecIV} we discuss the spin dynamics for boost-invariant system (Bjorken flow), the conservation of energy-momentum tensor, and the total angular momentum tensor for Bjorken flow. In Sec .~\ref {SecV} we present the numerical solution of the spin hydrodynamic equation for the boost-invariant system. Here, we show the proper time evolution of medium temperature and the spin chemical potential. In Sec.~\ref{SecVI}, we calculate the dilepton production using the solution of the dissipative spin hydrodynamic framework. We consider the proper time evolution of spin fluid to estimate the dilepton rate. Finally, in Sec.~\ref{SecVII} we conclude, discuss the limitations of the present investigation and possible future directions.

\section{Spin hydrodynamic framework}
\label{SecII}
Spin hydrodynamic frameworks are described by the following conservation laws~\cite{Florkowski:2017ruc},
\begin{align}
& \partial_{\mu}T^{\mu\nu}=0, ~~~~\partial_{\mu}J^{\mu}=0,~~~\partial_{\lambda}J^{\lambda\mu\nu}=0,\label{equ1ver1}
\end{align}
where,  $T^{\mu\nu}$ is the energy-momentum tensor which, in general, may not be symmetric under the exchange of $\mu\leftrightarrow\nu$, e.g., the canonical energy-momentum tensor obtained from Noether's theorem is not manifestly symmetric. In the above equation, $J^{\mu}$ is the global conserved current. Note that QCD constituents may carry multiple conserved charges, e.g., net baryon number $B$, net strangeness $S$, and electric charge $Q$. The total angular momentum tensor is represented by $J^{\mu\alpha\beta}$ which is the sum of the orbital angular momentum tensor $L^{\lambda\mu\nu}$, and the spin tensor $S^{\lambda\mu\nu}$~\cite{Florkowski:2018fap}. The orbital part can be expressed as $L^{\lambda\mu\nu}=x^{\mu}T^{\lambda\nu}-x^{\nu}T^{\lambda\mu}$. We observe that $L^{\lambda\mu\nu}$ is anti-symmetric in the last two indices, but the spin tensor $S^{\lambda\mu\nu}$ can be totally anti-symmetric, e.g., the canonical spin tensor. We emphasize that, in general, neither $L^{\lambda\mu\nu}$, nor $S^{\lambda\mu\nu}$ are separately conserved. The  conservation of the total angular momentum tensor implies the following condition for four-divergence of the spin tensor, 
\begin{align}
\partial_{\lambda}S^{\lambda\mu\nu}=-T^{\mu\nu}+T^{\nu\mu}=-2 T^{[\mu\nu]}.\label{equ3ver1}     
\end{align}
It is evident from the above equation that for a symmetric energy-momentum tensor ($T^{\mu\nu}=T^{\nu\mu}$), the spin tensor is separately conserved, i.e., $\partial_{\lambda}S^{\lambda\mu\nu}=0$. 
The anti-symmetric part of the energy-momentum tensor gives rise to the spin-orbit conversion, which spoils the separate conservation of the spin tensor.

In general $T^{\mu\nu}$, $J^{\mu}$, and $S^{\lambda\mu\nu}$ contains dissipative corrections. In standard hydrodynamics (spin-less fluid dynamics), in the absence of dissipation, $T^{\mu\nu}$ and $J^{\mu}$ can be expressed as, 
\begin{align}
& T^{\mu\nu}_{(0)}=\varepsilon u^{\mu} u^{\nu}-P\Delta^{\mu\nu}\label{equ5ver2}\\
& J^{\mu}_{(0)}=n u^{\mu}\label{equ6ver2}.
\end{align}
Here $\varepsilon$, $P$, and $n$ are energy-density, pressure, and number density. $\varepsilon$, $P$, and $n$ are not independent, but related by the equation-of-state (EoS), $P(\varepsilon,n)$. These quantities can be completely specified by temperature ($T$), and chemical potential ($\mu$). Hence, $T^{\mu\nu}_{(0)}$, and $J^{\mu}_{(0)}$ are completely specified by $T$, $\mu$, and $u^{\mu}$. 
Together $T$, $\mu$, and $u^{\mu}$ have five degrees of freedom. Dynamics of $T$, $\mu$, and $u^{\mu}$ is determined by the conservation equations $\partial_{\mu}T^{\mu\nu}_{(0)}=0$ and $\partial_{\mu}J^{\mu}_{(0)}=0$, i.e., five dynamical equations. In the spin hydrodynamic framework, six additional equations emerge from the conservation of the total angular momentum tensor, $\partial_{\lambda}J^{\lambda\mu\nu}_{(0)}=0$. These six equations determine the dynamics of another anti-symmetric tensor having six degrees of freedom. This anti-symmetric tensor is identified as the spin chemical potential ($\omega^{\mu\nu}$)~\cite{Florkowski:2018fap}. The evolution of the spin chemical potential encodes the dynamics of the spin tensor. Note that if the spin tensor $S^{\lambda\mu\nu}=0$ (unpolarized medium), then the conservation of the total angular momentum tensor does not give rise to new additional dynamical equations. In this case, the energy-momentum tensor is symmetric, and conservation of the energy-momentum tensor implies the conservation of the total angular momentum tensor. 

Incorporating dissipative effects, the energy-momentum tensor, conserved current, and the spin tensor can be expressed as a hydrodynamic gradient expansion~\cite{Kovtun:2019hdm},
\begin{align}
& T^{\mu\nu}=\mathcal{O}(1)+\mathcal{O}(\partial)+\mathcal{O}(\partial^2)+\cdots\label{equ7ver2}\\
& J^{\mu}=\mathcal{O}(1)+\mathcal{O}(\partial)+\mathcal{O}(\partial^2)+\cdots\label{equ8ver2}\\
& S^{\lambda\mu\nu} = \mathcal{O}(1)+\mathcal{O}(\partial)+\mathcal{O}(\partial^2)+\cdots\label{equ9ver2}
\end{align}
Here $\mathcal{O}(1)$ represent leading order term, and $\mathcal{O}(\partial^k)$ represent $k$-th order term in the hydrodynamic gradient expansion. In this paper we keep up to $\mathcal{O}(\partial)$ terms in the constitutive relations (Eqs.~\eqref{equ7ver2}-\eqref{equ9ver2}). Note that $T$, $\mu$, $u^{\mu}$ are all leading order term ($\mathcal{O}(1)$) in the hydrodynamic gradient expansion. What one needs to specify is the gradient ordering of the spin chemical potential ($\omega^{\alpha\beta}$). In literature, different hydrodynamic ordering of the spin chemical potential has been considered, e.g., in Refs.~\cite{Hattori:2019lfp,Biswas:2023qsw,Biswas:2022bht,Daher:2022xon,Daher:2022wzf} authors considered 
$\omega^{\mu\nu}\sim \mathcal{O}(\partial)$. It can be argued that, for asymmetric energy-momentum tensor (i.e., with antisymmetric parts), spin chemical potential $\omega^{\mu\nu}$ is completely determined by the thermal vorticity  $\varpi^{\mu\nu}\equiv -(\partial_{\mu}(u_{\nu}/T)-\partial_{\nu}(u_{\mu}/T))/2$ in global equilibrium~\cite{Becattini:2013fla,Florkowski:2018fap,Rindori:2020qqa}. This is the rationale behind the hydrodynamic gradient ordering of spin chemical potential, $\omega^{\mu\nu}\sim \mathcal{O}(\partial)$ ~\cite{Hattori:2019lfp,Biswas:2023qsw,Biswas:2022bht,Daher:2022xon,Daher:2022wzf}. However, if the energy-momentum tensor is symmetric, then in global equilibrium the spin chemical potential and the thermal vorticity need not be related~\cite{She:2021lhe,Dey:2024cwo}.  In that case one can consider $\omega^{\mu\nu}\sim \mathcal{O}(1)$ for theoretical consistency~\cite{She:2021lhe,Daher:2022wzf,Dey:2024cwo}. 
In this article, we consider that the energy-momentum tensor is symmetric and the spin hydrodynamic approach where $\omega^{\mu\nu}\sim \mathcal{O}(1)$ term in the gradient expansion.

Another nontrivial intricacy associated with spin hydrodynamic frameworks is the pseudo-gauge transformation~\cite{Chen:2018cts,HEHL197655,Speranza:2020ilk}. Pseudo-gauge transformation implies that, in the presence of an appropriate super-potential $\Phi^{\lambda,\mu\nu}$, we can redefine  $T^{\mu\nu}$, and 
$S^{\lambda,\mu\nu}$ without affecting the conservation of the energy-momentum tensor and the total angular momentum tensor. If $\partial_{\mu}T^{\mu\nu}=0$, and $\partial_{\lambda}J^{\lambda,\mu\nu}=0$, then we can also find $\partial_{\mu}T^{\prime\mu\nu}=0$, and $\partial_{\lambda}J^{\prime\lambda,\mu\nu}=0$, where $T^{\mu\nu}$, and $S^{\lambda\mu\nu}$ are related to the modified energy-momentum tensor $T^{\prime\mu\nu}$, and spin tensor $S^{\prime\lambda\mu\nu}$ in the following manner, $T^{\prime\mu\nu}=T^{\mu\nu}+\frac{1}{2}\partial_{\lambda}\left(\Phi^{\lambda,\mu\nu}-\Phi^{\mu,\lambda\nu}-\Phi^{\nu,\lambda\mu}\right)$, and $S^{\prime\lambda,\mu\nu}=S^{\lambda,\mu\nu}-\Phi^{\lambda,\mu\nu}$. Such transformations of $T^{\mu\nu}$, and $S^{\lambda\mu\nu}$ are known as the pseudo-gauge transformation~\cite{Speranza:2020ilk}. The super potential $\Phi^{\lambda,\mu\nu}$ is anti-symmetric at least in the last two indices~\cite{Florkowski:2018fap}. Different choices of $\Phi^{\lambda,\mu\nu}$
represents different pseudo-gauge transformations, e.g., Belinfante-Rosenfeld (BR) pseudo-gauge~\cite{BELINFANTE1939887,BELINFANTE1940449,Rosenfeld1940}, the de Groot-van Leeuwen-van  Weert  (GLW) pseudo-gauge~\cite{DeGroot:1980dk}, the Hilgevoord-Wouthuysen (HW) pseudo-gauge~\cite{HILGEVOORD19631,HILGEVOORD19651002}, etc.

Here we consider a symmetric energy-momentum tensor, and the spin tensor has a simple \textit{phenomenological} form. In the \textit{phenomenological} form, the spin tensor is only antisymmetric in the last two indices~\cite{Weyssenhoff:1947iua,Florkowski:2018fap,Florkowski:2017ruc}. Moreover, the leading order term, $S^{\lambda\mu\nu}_{(0)}$ can be expressed as, $S^{\lambda\mu\nu}_{(0)}=u^{\lambda}S^{\mu\nu}$. $S^{\mu\nu}$ is the spin density~\cite{Hattori:2019lfp,Daher:2022xon}. Such a choice of energy-momentum tensor and the spin tensor has been  used to develop a theoretically consistent spin hydrodynamic framework~\cite{Dey:2024cwo}. In this framework incorporating $\mathcal{O}(\partial)$ terms, the $T^{\mu\nu}$, $J^{\mu}$ and $S^{\lambda\mu\nu}$ can be expressed as~\cite{Dey:2024cwo,Biswas:2023qsw}, 
\begin{align}
    & T^{\mu\nu}=T^{\mu\nu}_{(0)}+T^{\mu\nu}_{(1)}\,\nonumber\\
    & \quad~~~ =\varepsilon u^\mu u^\nu -P \Delta^{\mu\nu}\nonumber\\
    & \quad~~~~~~~~ +h^{\mu} u^\nu+h^{\nu} u^{\mu} +\pi^{\mu\nu}+\Pi \Delta^{\mu\nu}, \label{equ10ver2}\\
    & J^{\mu}=J^{\mu}_{(0)}+ J^{\mu}_{(1)}=nu^{\mu}+ J^{\mu}_{(1)},\label{equ11ver2}\\
        & S^{\mu\alpha\beta}=S^{\mu\alpha\beta}_{(0)}+S^{\mu\alpha\beta}_{(1)}\nonumber\\ 
& \quad ~~~~ = u^\mu S^{\alpha\beta} \nonumber\\
& \quad ~~+ 2 u^{[\alpha} \Delta^{\mu\beta]} \Phi
+ 2 u^{[\alpha} \tau_{(s)}^{\mu\beta]}
+ 2 u^{[\alpha} \tau_{(a)}^{\mu\beta]}
+ \Theta^{\mu\alpha\beta}. \label{equ12ver2}
\end{align}
$\varepsilon$ is the energy density, $P$ is the pressure, $n$ is the number density, and $S^{\alpha\beta}$ is the spin density. $u^{\mu}$ is the normalized fluid velocity $u^{\mu}u_{\mu}=1$. $\pi^{\mu\nu}$ is a symmetric traceless tensor representing shear stress, $\Pi$ is the bulk viscous pressure and $h^{\mu}$ is the  energy diffusion four-current. $\pi^{\mu\nu}$, $\Pi$, and $h^{\mu}$ are all $\mathcal{O}(\partial)$ terms in the hydrodynamic gradient ordering. They satisfy the following conditions, $\pi^{\mu\nu}=\pi^{\nu\mu}$, $\pi^{\mu}_{~\mu}=0$, $\pi^{\mu\nu}u_{\mu}=0$, $h^{\mu}u_{\mu}=0$. The third rank tensor $S^{\mu\alpha\beta}_{(1)}$ which is antisymmetric in last two indices can be divided into a scalar ($\Phi$), a second rank symmetric tensor ($\tau^{\mu\nu}_{(s)}$), a second rank anti-symmetric tensor ($\tau^{\mu\nu}_{(a)}$), and a third rank tensor ($\Theta^{\mu\alpha\beta}$). $\Phi,\tau^{\mu\nu}_{(s)}, \tau^{\mu\nu}_{(a)} $, and $\Theta^{\mu\alpha\beta}$ are the $\mathcal{O}(\partial)$ terms that appear in the spin tensor~\cite{Biswas:2023qsw}. These currents satisfy the following conditions: $\tau_{(s)}^{\mu \nu}=\tau_{(s)}^{\nu \mu}$, $\tau_{(s)\mu}^{\mu}=0$, $\tau_{(s)}^{\mu \nu}u_{\nu}=0$, $\tau_{(a)}^{\mu \nu}=-\tau_{(a)}^{\nu \mu}$, $\tau_{(a)}^{\mu \nu}u_{\mu}=0$, $\Theta^{\mu \alpha \beta}=-\Theta^{\mu  \beta\alpha}$; $u_{\mu} \Theta^{\mu \alpha \beta}=0$; $u_{\alpha} \Theta^{\mu \alpha \beta}=0$. Note that $u_{\lambda}S^{\lambda\alpha\beta}_{(1)}=0$. 
$\Phi$ has 1 component,  $\tau^{\mu \beta}_{(s)}$ is symmetric traceless and orthogonal to $u^{\mu}$ having 5 components. Similarly, $\tau^{\mu \beta}_{(a)}$ has 3 components and $\Theta^{\mu\alpha\beta}$ is antisymmetric in last two indices and orthogonal to $u^{\mu}$ in all indices having 9 components giving us a total of 18 independent component as required for $S^{\mu\alpha\beta}_{(1)}$. Note that $S^{\mu\alpha\beta}$ has 24 components, and  six components of $S^{\mu\alpha\beta}_{(0)}$ stems from the six components of $S^{\alpha\beta}$~\cite{Biswas:2023qsw}. $J^{\mu}_{(1)}$ is the $\mathcal{O}(\partial)$ term, which satisfy the condition, $J^{\mu}_{(1)}u_{\mu}=0$.

The constitutive relations for different $\mathcal{O}(\partial)$ terms that appear in $T^{\mu\nu}$, $J^{\mu}$, and $S^{\lambda\mu\nu}$ can be obtained using the entropy current analysis, where we write down the entropy four current for the dissipative system~\cite{Hattori:2019lfp,Daher:2022xon,Biswas:2023qsw}, 
\begin{align}
\mathcal{S}^{\mu}=T^{\mu\nu}\beta_{\nu}+P\beta^{\mu}-\alpha J^{\mu}-\beta\omega_{\alpha\beta}S^{\mu\alpha\beta},
\label{equ13ver2}
\end{align}
here $\beta^{\mu}=\beta u^{\mu}=u^{\mu}/T$, and $\alpha=\mu/T$. Using Eqs.~\eqref{equ10ver2}-\eqref{equ12ver2}, back in to Eq.~\eqref{equ13ver2}, and demanding $\partial_{\mu}\mathcal{S}^{\mu}\geq0$, one can find the constitutive relations for $h^{\mu}$, $\pi^{\mu\nu}$, $\Pi$, $J^{\mu}_{(1)}$, $\Phi$, $\tau^{\mu\nu}_{(s)}$, $\tau^{\mu\nu}_{(a)}$, and $\Theta^{\mu\alpha\beta}$ in terms of the derivatives of $T$, $\mu$, $u^{\mu}$, and $\omega^{\mu\nu}$~\cite{Dey:2024cwo}. In the Landau frame $h^{\mu}=0$ \footnote{The general expression of $h^{\mu}$ can be shown to be: $h^{\mu}=-\kappa_{11}\frac{S^{\alpha\beta}}{\varepsilon+P}\nabla^{\mu}(\beta\omega_{\alpha\beta})-\kappa_{12}\nabla^{\mu}\alpha$~\cite{Dey:2024cwo}. For a baryon free system and for Bjorken flow $h^{\mu}$ identically vanishes. Naturally, one can apply the Landau frame condition.}, for the baryon free medium  \footnote{ In QCD, the net baryon number is conserved, i.e., the difference between the baryon number and the anti-baryon number is conserved. In our study, a baryon-free system implies that the number of baryons and anti-baryons is the same, i.e., the net baryon number is zero. This is realized by setting the baryon chemical potential ($\mu$) to zero. Hence, the net baryon number conservation, or loosely said, the baryon number conservation, is not violated. This assumption is commonly used to simplify the calculation in other references as well, e.g., Refs.~\cite{Biswas:2022bht,Wang:2021ngp,Wang:2024afv,Hattori:2019lfp}.}, the relevant dissipative currents are $\pi^{\mu\nu}$, $\Pi$, $\Phi$, $\tau^{\mu\nu}_{(s)}$, $\tau^{\mu\nu}_{(a)}$, and $\Theta^{\mu\alpha\beta}$. Their 
constitutive relations are given as~\cite{Dey:2024cwo}
\begin{align}
& \Pi=\zeta\theta,\label{equ14ver2}\\
& \pi^{\mu\nu}=2\eta\sigma^{\mu\nu},\label{equ15ver2}\\
& \Phi=-2\chi_{1} u^{\alpha}\nabla^{\beta}(\beta\omega_{\alpha\beta}),
\label{equ16ver2}\\
& \tau^{\mu\beta}_{(s)}=-2\chi_{2}\Delta^{\mu\beta,\gamma\rho}\nabla_{\gamma}(\beta\omega_{\alpha\rho})u^{\alpha},\label{equ17ver2}\\
& \tau^{\mu\beta}_{(a)}=-2\chi_{3} \Delta^{[\mu\beta][\gamma\rho]}\nabla_{\gamma}(\beta\omega_{\alpha\rho})u^{\alpha},
  \label{equ18ver2}\\
& \Theta^{\mu\alpha\beta}= \chi_4 \Delta^{\delta\alpha}\Delta^{\rho\beta}\Delta^{\gamma\mu}\nabla_{\gamma}(\beta\omega_{\delta\rho})\label{equ19ver2}.
\end{align}
$\eta$, $\zeta$, $\chi_1$, $\chi_2$, $\chi_3$, and $\chi_4$ are different transport coefficients. The positivity of the entropy production implies, 
$\eta\geq 0$, $\zeta\geq 0$, $\chi_1\geq 0$, $\chi_2\geq0$, $\chi_3\geq0$, and $\chi_4\geq0$. To complete the spin hydrodynamic framework, we also need the thermodynamic relations satisfied by the thermodynamic quantities. For the baryon-free system, these thermodynamic relations can be written as~\cite{Dey:2024cwo,Biswas:2022bht}, 
\begin{align} 
& \varepsilon+P =Ts+\omega_{\alpha\beta}S^{\alpha\beta},\label{equ20ver2}\\
& d\varepsilon =Tds+\omega_{\alpha\beta}dS^{\alpha\beta},\label{equ21ver2}\\ 
& dP=sdT+S^{\alpha\beta}d\omega_{\alpha\beta}.
\label{equ22ver2}
\end{align}
Here $s$ is the entropy density in local equilibrium. The above thermodynamic relations also imply, 
\begin{align}
s=\left.\frac{\partial P}{\partial T}\right|_{\omega^{\alpha\beta}}, ~~S^{\alpha\beta}=\left.\frac{\partial P}{\partial \omega_{\alpha\beta}}\right|_{T}
\label{equ23ver2}
\end{align}
In the baryon-free system, all thermodynamic quantities are functions of temperature ($T$) and spin chemical potential ($\omega^{\mu\nu}$). In general $P(T,\omega^{\mu\nu})$, $\varepsilon(T,\omega^{\mu\nu})$, and $s(T,\omega^{\mu\nu})$, can be obtained from a underlying microscopic theory. However, in the absence of such a microscopic theory, we can write $P(T,\omega^{\mu\nu})$ in the following way~\cite{Biswas:2022bht}, 
\begin{align}
    & P(T,\omega^{\mu\nu})=P_0(T)+ P_1(T)~ \omega^{\mu\nu}\omega_{\mu\nu}\label{equ24ver2}.
\end{align}
Here, $P_{0}(T)$ and $P_{1}(T)$ only depend on the temperature, and the second term includes the effect of the spin chemical potential\footnote{Here we have not incorporated higher order terms in $\omega^{\mu\nu}\omega_{\mu\nu}$, e.g., $(\omega^{\mu\nu}\omega_{\mu\nu})^2$, because we consider $\omega^{\mu\nu}/T$ small. This is a small polarization limit, where one considers the dimensionless ratio $\omega^{\mu\nu}/T<1$~\cite{Florkowski:2019qdp}.   }. Note that in our calculation both $T$, and $\omega^{\mu\nu}$ are leading order in the hydrodynamic gradient expansion, hence $P(T,\omega^{\mu\nu})\sim\mathcal{O}(1)$. Moreover, in the limit $\omega^{\alpha\beta}\rightarrow 0$, one obtains $P(T,\omega^{\mu\nu})=P_{0}(T)$, which is the pressure for the spin-less fluid.  
Using the expression of $P(T,\omega^{\mu\nu})$ in Eq.~\eqref{equ23ver2} one finds, 
\begin{align}
    & S^{\mu\nu} (T,\omega^{\mu\nu})=\left.\frac{\partial P}{\partial \omega_{\mu\nu}} \right|_{T} = S_0(T) \omega^{\mu \nu},
    \label{equ25ver2}
\end{align}
where $S_{0}(T)=2 P_1(T)$. The above relation between the spin density ($S^{\mu\nu}$), and the spin chemical potential ($\omega^{\mu\nu}$) is called the spin equation-of-state. Note that Eq.~\eqref{equ25ver2} is also consistent with the hydrodynamic gradient expansion. Both $S^{\mu\nu}$, and $\omega^{\mu\nu}$ are leading order terms, i.e., $\mathcal{O}(1)$ terms. Moreover, $P_{1}(T)$ is also a function of temperature and does not involve any derivatives. Note that in Ref.~\cite{Wang:2021ngp}, the authors considered a different hydrodynamic ordering of the spin chemical potential. They considered $\omega^{\mu\nu}\sim\mathcal{O}(\partial)$, and $S^{\mu\nu}\sim\mathcal{O}(1)$.
However, they have also considered that $S^{\mu\nu}\sim \omega^{\mu\nu}$. To establish such a relation in Ref.~\cite{Wang:2021ngp}, the authors consider that $S^{\mu\nu}\sim T^2\omega^{\mu\nu}$, and argued that only in the high temperature limit, a leading order ($\mathcal{O}(1)$) term should be related to the sub-leading ($\mathcal{O}(\partial)$) term. Considering the issue in connecting a term of the order of $\mathcal{O}(1)$ and a term of the order of $\mathcal{O}(\partial)$, in Ref.~\cite{Biswas:2022bht} authors propose an alternative form of the spin equation-of-state. They considered that, $S^{\mu\nu} (T,\omega^{\mu\nu}) = S_0(T) \omega^{\mu \nu}/\sqrt{\omega^{\mu\nu}\omega_{\mu\nu}}$. This relation is consistent with the hydrodynamic gradient ordering as the LHS and RHS of $S^{\mu\nu} (T,\omega^{\mu\nu}) = S_0(T) \omega^{\mu \nu}/\sqrt{\omega^{\mu\nu}\omega_{\mu\nu}}$  are both leading order ($\mathcal{O}(1)$). But for the spin equation-of-state $S^{\mu\nu} (T,\omega^{\mu\nu}) = S_0(T) \omega^{\mu \nu}/\sqrt{\omega^{\mu\nu}\omega_{\mu\nu}}$, one can not simply consider the $\omega^{\mu\nu}\rightarrow 0$ limit, to obtain the standard hydrodynamics (spin-less fluid dynamics) from the spin hydrodynamic framework. The spin equation-of-state that we use (Eq.~\eqref{equ25ver2}) is free of such conceptual issues, it is consistent with the hydrodynamic gradient expansion, and one can also consider the $\omega^{\mu\nu}\rightarrow 0$ limit to obtain the standard hydrodynamic framework as a limiting case.

Now using Eq.~\eqref{equ24ver2}, in Eq.~\eqref{equ23ver2} we find the expression of $s(T,\omega^{\mu\nu})$, 
\begin{align}
 s(T, \omega^{\mu\nu})=\left. \frac{\partial P}{\partial T} \right|_{\omega_{\mu\nu}} = s_0(T) + \frac{1}{2} S_0'(T)~\omega^{\mu \nu} \omega_{\mu\nu}\label{equ26ver2}
\end{align}
here, $s_0(T)\equiv dP_0(T)/dT$, and  $S_0'(T)\equiv dS_0/dT$.  Using Eqs.~\eqref{equ24ver2}-\eqref{equ26ver2}, back into Eq.~\eqref{equ20ver2} we find, 
\begin{align}
\varepsilon(T, \omega^{\mu\nu})= \varepsilon_0(T) + \frac{1}{2} \bigg[S_0(T) + T S_0'(T)\bigg]\omega^{\mu\nu}\omega_{\mu\nu},\label{equ27ver2}
\end{align}
here $\varepsilon_0(T)+P_{0}(T)=Ts_0(T)$ is the first law of thermodynamics for spin-less fluid. 

We have the constitutive relations for the dissipative currents, along with the thermodynamic relations and the spin equation-of-state. Now we can write the spin hydrodynamic equations for the baryon-free system in the Landau frame,  
\begin{align}
    & u^{\mu} \partial_{\mu}\varepsilon+(\varepsilon+P-\Pi) \theta -\pi^{\mu\nu}\partial_{\mu}u_{\nu}=0, \label{equ28ver2}\\ 
    & \left(\varepsilon+P-\Pi\right)\left(u^{\mu} \partial_{\mu}\right) u^{\alpha}-\Delta^{\alpha\mu}\partial_{\mu} \left(P-\Pi\right) \nonumber\\
    & ~~~~~~~~~~~~~~~~~~~~~~~~~~~+\Delta^{\alpha}_{~~\nu} \partial_{\mu} \pi^{\mu\nu}=0, \label{equ29ver2}\\ 
    & u^{\mu}\partial_\mu S^{\alpha\beta}+S^{\alpha\beta}\partial_{\mu} u^{\mu}+\partial_{\mu}S^{\mu\alpha\beta}_{(1)}=0.\label{equ30ver2}
\end{align}
Eq.~\eqref{equ28ver2} is the projection of $\partial_{\mu}T^{\mu\nu}=0$ along the direction of $u^{\mu}$. Eq.~\eqref{equ29ver2} is the projection of $\partial_{\mu}T^{\mu\nu}=0$ normal to the the direction of $u^{\mu}$, i.e., $\Delta^{\alpha}_{~~\nu}\partial_\mu T^{\mu\nu}=0$. The third equation (Eq.~\eqref{equ30ver2}) is nothing but the conservation of the total angular momentum tensor. Note that in our calculation, we consider a symmetric energy-momentum tensor, hence $\partial_{\mu} J^{\mu\alpha\beta}=0$ implies the conservation of the spin tensor, $\partial_{\mu} S^{\mu\alpha\beta}=0$. To solve Eqs.~\eqref{equ28ver2}-\eqref{equ30ver2}, we need to specify the fluid flow configuration. Here, we consider the boost-invariant flow, also known as the Bjorken flow~\cite{Bjorken:1982qr,Wang:2021ngp,Biswas:2022bht}.

\section{Spin dynamics in a boost invariant system}
\label{SecIII}
 In heavy-ion collisions, for the Bjorken flow~\cite{Bjorken:1982qr}, one assumes that the system expands along the longitudinal direction (beam direction). Moreover, the transverse direction (compared to the beam direction) is uniform, and there is no expansion in the transverse plane. Assuming the beam axis along the $Z$ direction, the fluid flow for the Bjorken flow in the Cartesian coordinate can be expressed as, $u^{\mu} = (\cosh \eta_s, 0, 0, \sinh \eta_s)$. Here $\eta_s =(1/2)\ln[(t+z)/(t-z)]$ is the spacetime rapidity. For Bjorken flow, scalar quantities, e.g., energy-density, pressure, number density, temperature, etc. only depend on the proper time $\tau=\sqrt{t^2-z^2}$. Cartesian coordinate $(t,x,y,z)$, can be expressed in terms of $(\tau,x,y,\eta_s)$ as, $t=\tau \cosh \eta_s$, $z=\tau \sinh \eta_s$.
For a boost-invariant system, to represent various dissipative currents, it is convenient to introduce the following set of vectors~\cite{Biswas:2022bht}:
\begin{align}
 u^{\mu}&\equiv\left(\cosh\eta_s,0,0,\sinh\eta_s\right),\label{equ31ver2}\\
 X^{\mu}&\equiv \left(0,1,0,0\right),\label{equ32ver2}\\
 Y^{\mu}&\equiv \left(0,0,1,0\right),\label{equ33ver2}\\
 Z^{\mu}&\equiv \left(\sinh\eta_s,0,0,\cosh\eta_s\right).\label{equ34ver2}
\end{align}
$u^{\mu}$ is a time-like four vector which satisfies $u^{\mu}u_{\mu}=1$. 
$X^{\mu}$ , $Y^{\mu}$, and $Z^{\mu}$ are space-like four vectors satisfying the conditions, $X_{\mu}X^{\mu}=-1$, $Y_{\mu}Y^{\mu}=-1$, $Z_{\mu}Z^{\mu}=-1$. It is evident from Eqs.~\eqref{equ31ver2}-\eqref{equ34ver2} that $u_{\mu}X^{\mu}=0$, $u_{\mu}Y^{\mu}=0$, $u_{\mu}Z^{\mu}=0$, $X_{\mu}Y^{\mu}=0$, $X_{\mu}Z^{\mu}=0$, and $Y_{\mu}Z^{\mu}=0$. Hence the set of vectors $\left(u^{\mu}, X^{\mu}, Y^{\mu}, Z^{\mu}\right)$ forms a orthonormal basis vectors.

The spin chemical potential ($\omega^{\mu\nu}$), being a second rank anti-symmetric tensor, can be decomposed into an electric-like component ($\kappa^{\mu}$), a magnetic-like component ($\omega^{\mu}$)~\cite{Florkowski:2018fap,Florkowski:2018ahw, Biswas:2022bht}
\begin{align}
 \omega^{\mu\nu}=\kappa^{\mu}u^{\nu}-\kappa^{\nu}u^{\mu}+\epsilon^{\mu\nu\alpha\beta}u_{\alpha}\omega_{\beta}.\label{equ35ver2}
\end{align}
Eq.~\eqref{equ35ver2} can be inverted to to write $\kappa^{\mu}$, and $\omega^{\mu}$ in terms of $\omega^{\mu\nu}$. Moreover it can be shown that $\kappa^{\mu}$, and $\omega^{\mu}$ are space-like, i.e., $\kappa^{\mu}u_{\mu}=0$, and $\omega^{\mu}u_{\mu}=0$. $\kappa^{\mu}$ and $\omega^{\mu}$ both have three independent components, which add up to the six independent components of the spin chemical potential. The space-like four vectors $\kappa^{\mu}$, and $\omega^{\mu}$ can expressed in term of $X^{\mu}$, $Y^{\mu}$, and $Z^{\mu}$,
\begin{align}
    \kappa^{\mu} & =C_{\kappa X}X^{\mu}+C_{\kappa Y}Y^{\mu}+C_{\kappa Z}Z^{\mu}\nonumber\\
    & = \left(C_{\kappa Z}\sinh\eta_s,C_{\kappa X},C_{\kappa Y},C_{\kappa Z}\cosh\eta_s\right)\\
    \omega^{\mu} & = C_{\omega X}X^{\mu}+C_{\omega Y}Y^{\mu}+C_{\omega Z}Z^{\mu}\nonumber\\
    & =\left(C_{\omega Z}\sinh\eta_s,C_{\omega X},C_{\omega Y},C_{\omega Z}\cosh\eta_s\right).
\end{align}
The coefficients $\left(C_{\kappa X}, C_{\kappa Y}, C_{\kappa Z}\right)$, and $\left(C_{\omega X}, C_{\omega Y}, C_{\omega Z}\right)$ only depend on the proper time ($\tau$). Some of these coefficients are not relevant to determining the spin dynamics in a boost-invariant system. Physical implications of these coefficients, $C_{\kappa i}$, and $C_{\omega i}$ ($i\in (X,Y,Z)$) can be understood 
by looking into the total angular momentum of the fire-cylinder (FC) defined by the conditions: $\tau$=constant, $-\eta_{\rm FC}/2 \leq \eta_s \leq \eta_{\rm FC}/2$, and $\sqrt{x^{2}+y^{2}}\leq R$ (see also Fig.~1 in Ref.~\cite{Florkowski:2019qdp} and the discussion given in Ref.~\cite{Biswas:2022bht}). $R$ is the transverse size of the system. The total angular momentum of the fire-cylinder (FC) is defined as, 
\begin{align}
\mathcal{J}^{\mu\nu}_{\rm FC}=\mathcal{L}^{\mu\nu}_{\rm FC}+\mathcal{S}^{\mu\nu}_{\rm FC}, 
\end{align}
here $\mathcal{L}^{\mu\nu}_{\rm FC}$, and $\mathcal{S}^{\mu\nu}_{\rm FC}$ are the orbital angular momentum and the spin angular momentum, respectively. $\mathcal{L}^{\mu\nu}_{\rm FC}$, and $\mathcal{S}^{\mu\nu}_{\rm FC}$ are defined as, 
\begin{align}
&     \mathcal{L}^{\mu\nu}_{\rm FC}=\int_{}^{} d\Sigma_{\lambda}L^{\lambda\mu\nu}_{};\quad 
 \mathcal{S}^{\mu\nu}_{\rm FC}=\int_{}^{} d\Sigma_{\lambda}S^{\lambda\mu\nu}_{}. 
    \label{equ39ver2}
\end{align}
$L^{\lambda\mu\nu}$ is the orbital angular momentum tensor, and $S^{\lambda\mu\nu}$ is the spin angular momentum tensor. $d\Sigma_{\lambda}\equiv u_{\lambda}dxdy\tau d\eta_s$ is the infinitesimal volume element of the fire-cylinder.  
Using the explicit form for the energy-momentum tensor (in the Landau frame) and the spin tensor, we obtain, 
\begin{align}
    & \mathcal{L}^{\mu\nu}_{\mathrm{FC}}
= \int d\Sigma_{\lambda}\, L^{\lambda\mu\nu}_{}
= \int d\Sigma_{\lambda}\,\left( x^{\mu} T^{\lambda\nu}_{}
- x^{\nu} T^{\lambda\mu}_{} \right)\nonumber\\
& ~~~~~~= \int dx\, dy\, d\eta_s~\tau \varepsilon \left( x^{\mu}u^{\nu}- x^{\nu}u^{\mu}\right)=0. \label{equ40ver2}\\
& \mathcal{S}^{\mu\nu}_{\mathrm{FC}}
=  \int d\Sigma_{\lambda}\, S^{\lambda\mu\nu}_{} =  \int d\Sigma_{\lambda}\, S^{\lambda\mu\nu}_{(0)} \nonumber\\
& ~~~~~= \int dx\, dy\, d\eta_s \, \tau \; S^{\mu\nu}_{}\nonumber\\
& ~~~~~= \int dx\, dy\, d\eta_s\, \tau\, 
S_{0}(T)\nonumber\\
& ~~~~~~~~~~~~~~~\times \left( \kappa^{\mu} u^{\nu} - \kappa^{\nu} u^{\mu}
+ \epsilon^{\mu\nu\alpha\beta} u_{\alpha} \omega_{\beta} \right). \label{equ41ver2}
\end{align}
In the last line of Eq.~\eqref{equ40ver2}, one uses the explicit form of Bjorken flow. $\mathcal{J}^{0i}_{\rm FC}$ (for $i=1,2,3$) components of the fire-cylinder describe its center-of-mass motion, and these components vanishes in the center-of-mass system. $\mathcal{J}^{0i}_{\rm FC}=0$ (for $i=1,2,3$) also implies that 
$\mathcal{S}^{0i}_{\rm FC}=0$ (for $i=1,2,3$). Using Eq.~\eqref{equ41ver2}  $\mathcal{S}^{0i}_{\rm FC}$ (for $i=1,2,3$) can be expressed as~\cite{Biswas:2022bht}, 
\begin{align}
\mathcal{S}^{01}_{\mathrm{FC}} &= - 2\pi R^{2}\tau S_{0} \, C_{\kappa X} 
\sinh\left( \frac{\eta_{\mathrm{FC}}}{2} \right), \\
\mathcal{S}^{02}_{\mathrm{FC}} &= - 2\pi R^{2}\tau S_{0} C_{\kappa Y} 
\sinh\left( \frac{\eta_{\mathrm{FC}}}{2} \right)\\
\mathcal{S}^{03}_{\mathrm{FC}} &= - \pi R^{2}\tau S_{0} C_{\kappa Z} \eta_{\mathrm{FC}}. 
\end{align}
The condition $\mathcal{S}^{0i}_{\rm FC}=0$ (for $i=1,2,3$) implies $C_{\kappa X} =0$, $C_{\kappa Y} =0$, and $C_{\kappa Z} =0$, i.e., $\kappa^{\mu}=0$.  Therefore, for the boost invariant system, the electric-like components ($\kappa^{\mu}$) of the spin chemical potential vanish in the center-of-mass system, and only magnetic-like components ($\omega^{\mu}$) survive \footnote{In Ref.~\cite{Daher:2024ixz} the authors proposed a general form of the spin equation-of-state, $S^{\mu\nu}=S_1 \left(\kappa^{\mu}u^{\nu}-\kappa^{\nu}u^{\mu}\right)+S_2\epsilon^{\mu\nu\alpha\beta}u_{\alpha}\omega_{\beta}$. When $S_1=S_2=S_0$, the general form of the spin equation-of-state boils down to Eq.~\eqref{equ25ver2}. Using this general form of the spin equation-of-state in Eq.~\eqref{equ41ver2} one can calculate the spin tensor of the fire cylinder ($\mathcal{S}^{\mu\nu}_{\rm FC}$). Moreover setting the condition, $\mathcal{S}^{0i}_{\rm FC}=0$ one can also argue that  $\kappa^{\mu}=0$. Hence, only the magnetic-like components are relevant for the boost-invariant system.}. In this case $\omega^{\mu\nu}=\epsilon^{\mu\nu\alpha\beta}u_{\alpha}\omega_{\beta}.$ which has the following independent components,  
\begin{align}
    & \omega^{01}=-C_{\omega Y} \sinh\eta_s; ~\omega^{03}=0,\\
    & \omega^{02}=C_{\omega X} \sinh\eta_s;\\
    & \omega^{13}=C_{\omega Y} \cosh\eta_s;~\omega^{12}=-C_{\omega Z},\\\
    & \omega^{23}=-C_{\omega X} \cosh\eta_s.
\end{align}
Moreover, it can be shown that, 
\begin{align}
\omega^{\mu\nu}\omega_{\mu\nu}=2 \left(C_{\omega X}^2+C_{\omega Y}^2+C_{\omega Z}^2\right)=2 C^2.\label{equ49ver2}
\end{align}
Using Eq.~\eqref{equ49ver2} in Eqs.~\eqref{equ24ver2}, \eqref{equ26ver2}, and \eqref{equ27ver2} the equilibrium thermodynamic quantities can be expressed as, 
\begin{align}
    & P(T,\omega^{\mu\nu})=P_0(T)+ S_0(T)~C^2,\label{equ50ver2}\\
    & \varepsilon(T, \omega^{\mu\nu})= \varepsilon_0(T) +  \bigg[S_0(T) + T S_0'(T)\bigg]C^2,\label{equ51ver2}\\
    & s(T, \omega^{\mu\nu})= s_0(T) + S_0'(T) C^2.\label{equ52ver2}
\end{align}

\section{Spin hydrodynamic equations for a boost-invariant system}
\label{SecIV}
For a boost invariant system, Eq.~\eqref{equ29ver2} is trivially satisfied (see Appendix.~\ref{appendix1} for a detailed derivation). On the other hand, Eq.~\eqref{equ28ver2} give rise to the proper time evolution of the energy density, 
\begin{align}
       \frac{d\varepsilon}{d\tau}+\frac{\varepsilon+P}{\tau}-\frac{s_0}{\tau^2}\bigg(\frac{4}{3}\frac{\eta}{s_0}+\frac{\zeta}{s_0}\bigg)=0.
    \label{equ53ver2}
\end{align}
Note that in general $\eta$, $\zeta$, $s_0$ all depend on the proper time ($\tau$). We write Eq.~\eqref{equ53ver2} in terms of the ratios $\eta/s_0$, and $\zeta/s_0$ as these are dimensionless quantities. Moreover, for a boost invariant system, Eq.~\eqref{equ30ver2} simplifies to, 
\begin{align}
    \frac{\partial S^{\alpha \beta}}{\partial \tau} +\frac{S^{\alpha \beta}}{\tau}+ \partial_\mu S^{\mu\alpha\beta}_{(1)}=0.
    \label{equ54ver2}
\end{align}
For the boost invariant system $S^{\mu\nu}=S_{0}(T)\epsilon^{\mu\nu\alpha\beta}u_{\alpha}\omega_{\beta}$.
To simplify Eq.~\eqref{equ54ver2}, we need explicit expressions for the different dissipative currents in the spin tensor. For Bjorken flow, it can be shown that, 
\begin{align}
& \Phi =0;~~~ \Theta^{\mu\alpha\beta}=0; \label{equ55ver2}\\
& \tau^{01}_{(a)}= -\chi_3\,\beta\,C_{\omega Y}\frac{\sinh\eta_s}{\tau};\tau^{02}_{(a)}= \chi_3\,\beta\,C_{\omega X}\frac{\sinh\eta_s}{\tau};\label{equ56ver2}\\
& \tau^{03}_{(a)}= 0;~\tau^{12}_{(a)}= 0; \label{equ57ver2}\\
& \tau^{13}_{(a)}= \chi_3\,\beta\,C_{\omega Y}\frac{\cosh\eta_s}{\tau};
\tau^{23}_{(a)}= -\chi_3\,\beta\,C_{\omega X}\frac{\cosh\eta_s}{\tau};\label{equ58ver2} \\
& \tau^{00}_{(s)}= 0;~\tau^{11}_{(s)}= 0;~\tau^{22}_{(s)}= 0;~\tau^{33}_{(s)}= 0; \label{equ59ver2}\\
& \tau^{01}_{(s)}= -\chi_2\,\beta\,C_{\omega Y}\frac{\sinh\eta_s}{\tau}; \tau^{02}_{(s)}= \chi_2\,\beta\,C_{\omega X}\frac{\sinh\eta_s}{\tau};\label{equ60ver2}\\
& \tau^{03}_{(s)}= 0;~\tau^{12}_{(s)}= 0;~ \tau^{13}_{(s)}= -\chi_2\,\beta\,C_{\omega Y}\frac{\cosh\eta_s}{\tau};\label{equ61ver2}\\
& \tau^{23}_{(s)}= \chi_2\,\beta\,C_{\omega X}\frac{\cosh\eta_s}{\tau}; \label{equ62ver2}
\end{align}
In general, Eq.~\eqref{equ54ver2} gives rise to six independent equations. These equations can be obtained by contracting Eq.~\eqref{equ54ver2} with $X_{\alpha}u_{\beta}$, $Y_{\alpha}u_{\beta}$, $Z_{\alpha}u_{\beta}$, $X_{\alpha}Y_{\beta}$, $X_{\alpha}Z_{\beta}$, and $Y_{\alpha}Z_{\beta}$. No non-trivial equations are obtained when we contract Eq.~\eqref{equ54ver2} with $X_{\alpha}u_{\beta}$, $Y_{\alpha}u_{\beta}$, and $Z_{\alpha}u_{\beta}$. However, when we contract Eq.~\eqref{equ54ver2} with $Y_{\alpha}Z_{\beta}$, $X_{\alpha}Z_{\beta}$, and $X_{\alpha}Y_{\beta}$ we find the proper time evolution of $C_{\omega X}$, $C_{\omega Y}$, and $C_{\omega Z}$, respectively (see Appendix~\ref{appendix2} for details),    
\begin{align}
    & \frac{d C_{\omega X}}{d \tau} +C_{\omega X} \bigg(\frac{S_0'(T)}{S_0(T)}\frac{dT}{d\tau}+\frac{1}{\tau}+ \frac{(\chi_2+\chi_3)}{\tau^2 T S_0(T)}\bigg)=0,\label{equ63ver2}\\
   & \frac{d C_{\omega Y}}{d \tau} +C_{\omega Y} \bigg(\frac{S_0'(T)}{S_0(T)}\frac{dT}{d\tau}+\frac{1}{\tau}+ \frac{(\chi_2+\chi_3)}{\tau^2 T S_0(T)}\bigg)=0,\label{equ64ver2}\\
    & \frac{d C_{\omega Z}}{d \tau} +C_{\omega Z} \bigg(\frac{S_0'(T)}{S_0(T)}\frac{dT}{d\tau}+\frac{1}{\tau}\bigg)=0.\label{equ65ver2}
\end{align}
Using Eqs.~\eqref{equ50ver2}-\eqref{equ51ver2}, and Eqs.~\eqref{equ63ver2}-\eqref{equ65ver2} in Eq.~\eqref{equ53ver2} one find the proper time evolution of temperature (see Appendix~\ref{appendix3} for details), 
\begin{align}
& \bigg[A-BC^2\frac{S_0'(T)}{S_0(T)}\bigg] \frac{dT}{d\tau}-\frac{BC^2}{\tau}-\frac{B\left(C_{\omega X}^2+C_{\omega Y}^2\right)}{T\tau^2}\chi_s\nonumber\\
& ~~~~+\frac{Ts_0(T)}{\tau}+\bigg[2S_0(T)+TS_0^{\prime}(T)\bigg]\frac{C^2}{\tau}\nonumber\\
& ~~~~~~~~~~~~~~~~~~~~~~~~~~~~-\frac{s_0}{\tau^2}\bigg(\frac{4\eta}{3 s_0}+\frac{\zeta}{s_0}\bigg)=0.
\label{equ66ver2}
\end{align}
Here, $\chi_s\equiv (\chi_2+\chi_3)/S_0(T)$, is the dimensionless ratio. $A$ and $B$ are defined as,
\begin{align}
    &A= \frac{d \varepsilon_0}{dT}+\bigg[2S_0'(T)+TS_0'' (T)\bigg] C^2,  \\ 
    &B=2\bigg[S_0(T)+TS_0'(T)\bigg].
\end{align}
Eqs.~\eqref{equ63ver2}-\eqref{equ66ver2} are coupled first-order differential equations which can be solved to obtain the proper time evolution of $T(\tau)$, $C_{\omega X}(\tau)$, $C_{\omega Y}(\tau)$, and $C_{\omega Z}(\tau)$. Considering the complexity of these equations, we solve Eqs.~\eqref{equ63ver2}-\eqref{equ66ver2} numerically to obtain the numerical solution of the dissipative spin hydrodynamic framework for Bjorken flow.

\section{Numerical solution of the spin hydrodynamic equation for a boost-invariant system}
\label{SecV}
To numerically solve Eqs.~\eqref{equ63ver2}-\eqref{equ66ver2} we need to specify $\varepsilon_0(T)$, $s_0(T)$, $S_0(T)$, $\chi_s$, $\eta/s_0$, and $\zeta/s_0$. We consider the energy density ($\varepsilon_0$) and the entropy density ($s_0$) of the medium as the energy density and entropy density of a non-interacting system of massive particles, having mass $m_0$ at temperature $T$~\cite{Biswas:2022bht},
\begin{align}
    & \varepsilon_0(T) = \frac{g_s\, m_0^{2}\, T^{2}}{2\pi^2}
\left[
3\ K_2\left(\frac{m_0}{T}\right)
+ \frac{m_0}{T}\, K_1\!\left(\frac{m_0}{T}\right)
\right]\\
    & s_0(T) = \frac{g_s\, m_0^3}{2\pi^2}\, K_3\!\left(\frac{m_0}{T}\right).
\end{align}
Here $K_n(m_0/T)$ is the modified Bessel function of the second kind of order $n$, $g_s=4$ is the spin and particle-antiparticle degeneracy factor (spin half particle). $\varepsilon_0(T)$, and $s_0(T)$ can also be used to obtain $P_0(T)$ using the thermodynamic relation $\varepsilon_0+P_0=Ts_0$. Note that $S_0(T)$ that appears in the spin equation-of-state can, in principle, be determined by a suitable microscopic calculation. However, in the absence of such a microscopic description, we treat $S_0(T)$ as a free parameter, and on dimensional grounds we consider $S_0(T)=s_0(T)/T$. This expression of $S_0(T)$ should be considered as a phenomenological ansatz ~\cite{Biswas:2022bht}.  Apart from various thermodynamic quantities, different transport coefficients $\eta$, $\zeta$, $\chi_2$, and $\chi_3$ enter in Eqs.~\eqref{equ63ver2}-\eqref{equ66ver2}. In general, transport coefficients are not constant, and they can change as the system evolves. The temperature dependence of these transport coefficients can be obtained either from a microscopic theory within the kinetic theory framework or from a quantum field theory approach within the Green-Kubo framework~\cite {FReif:1965,Green:1954,Kubo:1957mj}. Although such microscopic calculations to obtain $\eta$, and $\zeta$ are well developed for the standard hydrodynamic approach (spin-less fluid dynamics), estimation of spin transport coefficients using a microscopic calculation is still at the development stage~\cite{Hu:2021lnx,Dey:2024cwo,She:2024rnx,Daher:2025pfq}. For simplicity and qualitative estimation, in the present calculation, we consider fixed values of $\eta/s_0$. We ignore the effects of bulk viscosity $\zeta/s_0  \sim 0 $. Note that $\eta$, $\zeta$, and $s_0$ all depend on the proper time, but we assume that the dimensionless ratios $\eta/s_0$ and $\zeta/s_0$ remain constant. This is an approximation which has been considered in other references as well~\cite{Biswas:2022bht,Wang:2021ngp,Romatschke:2007mq}. Around the QCD critical point, the variation of $\eta/s_0$ and $\zeta/s_0$ is very important, but in the present calculation, such a situation does not appear, as we consider a baryon-free system.
For the spin transport coefficients, we also consider the dimensionless ratio $\chi_s=(\chi_2+\chi_3)/{S_0(T)}$ to be constant. 

\begin{figure}
    \centering
    \includegraphics[scale=0.55]{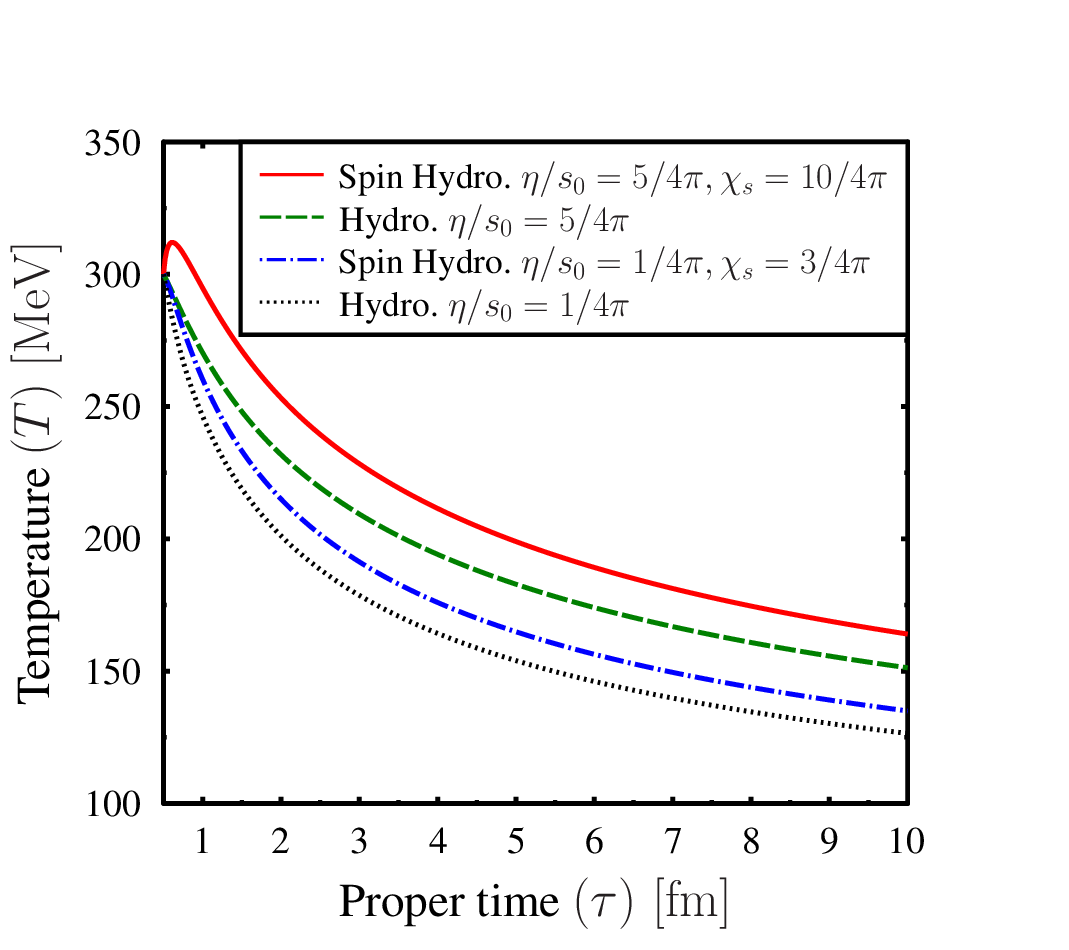}
    \caption{Proper time evolution of the medium temperature $T(\tau)$. The red solid line and the blue dashed-dotted line represent the temperature evolution in dissipative spin hydrodynamics. For this case, we consider $C_{\omega X}(\tau_0)=C_{\omega Y}(\tau_0)=C_{\omega Z}(\tau_0)=80$ MeV.  The green dashed line and the black dotted lines represent the temperature evolution for standard dissipative hydrodynamics. This is obtained by setting $C_{\omega X}=C_{\omega Y}=C_{\omega Z}=0$. In the limit $C_{\omega X}=C_{\omega Y}=C_{\omega Z}=0$ the spin hydrodynamic framework boils down to the standard dissipative hydrodynamics.}
    \label{fig:1}
\end{figure}

In Fig.~\ref{fig:1} we show the proper time evolution of the medium temperature $T(\tau)$ for dissipative spin hydrodynamics (solution of Eq.~\eqref{equ66ver2}). We choose the thermalization time scale $\tau_0 = 0.5$ fm to start the hydrodynamic evolution. At the initial time $\tau_0$ the initial temperature is $T(\tau_0)=T_0=300$ MeV and 
$C_{\omega X}(\tau_0)=C_{\omega Y}(\tau_0)=C_{\omega Z}(\tau_0)=80$ MeV.
We choose such a value of $C_{\omega X}$, $C_{\omega Y}$, and $C_{\omega Z}$ so that they are small as compared to the temperature. The mass of the medium particles is $m_0= 200$ MeV \footnote{Instead of considering mass-less particles, we consider a mass of 200 MeV, because in different references where authors compare theoretical predictions with spin polarization observables, a quasiparticle picture, with medium-dependent mass of similar, order becomes crucial~\cite{Fu:2021pok,Sapna:2025yss}.}. We consider two different choices of the transport coefficients, the first choice is $\eta/s_0=5/4\pi$, $\chi_s=10/4\pi$, and the second choice is $\eta/s_0=1/4\pi$, $\chi_s=3/4\pi$. For comparison, we also show the proper time evolution of temperature for standard dissipative hydrodynamics (spin-less fluid dynamics). Standard dissipative hydrodynamics (spin-less fluid dynamics) can be recovered from Eq.~\eqref{equ66ver2}) by setting $C_{\omega X}(\tau)=C_{\omega Y}(\tau)=C_{\omega Z}(\tau)=0$. In Fig.~\ref{fig:1}, the green dashed line and the black dotted line represent the temperature evolution for standard dissipative hydrodynamics for $\eta/s_0=5/4\pi$, and $\eta/s_0=1/4\pi$, respectively. On the other hand, the red solid line and the blue dashed dotted line represent the temperature evolution for dissipative spin hydrodynamics for  $\eta/s_0=5/4\pi$, $\chi_s=10/4\pi$, and $\eta/s_0=1/4\pi$, $\chi_s=3/4\pi$, respectively. From this figure, we observe that non vanishing spin chemical potential does affect the temperature evolution. The slower cooling observed in our calculations originates from the coupling between the spin sector and the thermodynamic evolution through the spin-dependent equation of state, Eqs.~\eqref{equ50ver2}-\eqref{equ52ver2}, and the coupled evolution equations Eqs.~\eqref{equ63ver2}-\eqref{equ66ver2}. The additional spin contribution acts as an extra reservoir of energy and entropy, leading to a modest reduction of the cooling rate compared to the corresponding spinless Bjorken evolution. Dissipative effects also slow the decrease in temperature over time. Moreover, we observe that in spin hydrodynamics for $\eta/s_0=5/4\pi$ and $\chi_s=10/4\pi$, the temperature initially increases for a brief moment, then decreases with proper time. This is in analogy with the reheating effect observed in the conventional first-order dissipative hydrodynamics where the solutions of the relativistic Navier-Stokes equation shows an unphysical reheating~\cite{Baier:2006um}. This is attributed to the fact that at early times, the gradients are quite large as they are governed by the factor $1/\tau$, leading to large entropy density increase per unit time. Therefore, for these early times, dissipative effects are significantly large and lies outside the validity of the relativistic Navier-Stokes equations, leading to the unphysical reheating. We note that, for sufficiently large values of $\chi_{s}$, the system exhibits a brief initial reheating stage. However, since $\chi_{s}$ is not determined from first principles in the present study, a quantitative estimate of the spin transport coefficients is required to place physical bounds on its magnitude. Such an analysis would clarify whether this reheating behavior can arise in realistic systems.

\begin{figure}
    \includegraphics[scale=0.55]{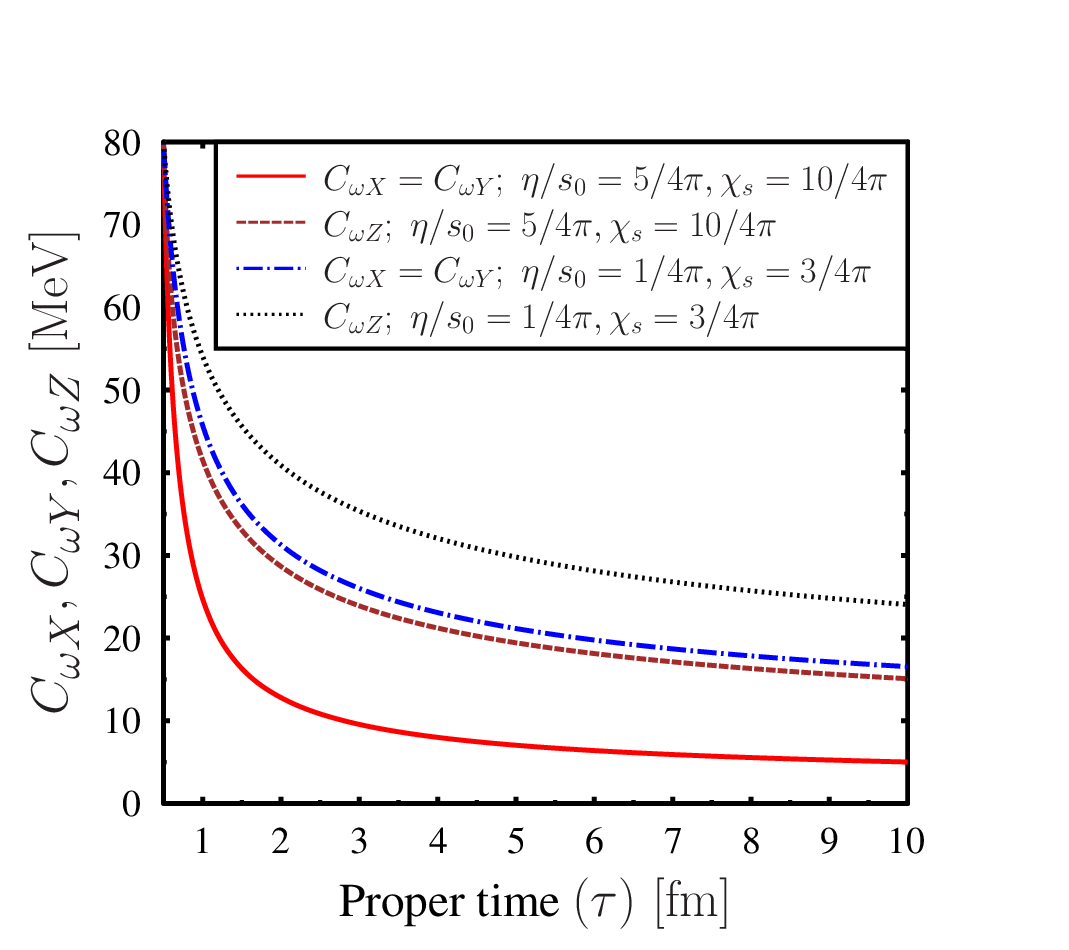}
    \caption{We show the evolution of different components of the spin chemical potential, i.e., $C_{\omega X}(\tau)$, $C_{\omega Y}(\tau)$, and $C_{\omega Z}(\tau)$. We observe that with proper time $C_{\omega X}(\tau)$, $C_{\omega Y}(\tau)$, and $C_{\omega Z}(\tau)$ decreases. The decrease of $C_{\omega X}(\tau)$, and $C_{\omega Y}(\tau)$ is faster as comped to $C_{\omega Z}(\tau)$, due to the spin dissipation. Note $\eta/s_0$ does not directly affect the evolution of spin chemical potential. But $\eta/s_0$ affects the temperature evolution, thereby indirectly affecting the evolution of the spin chemical potential. Due to the symmetry in the transverse plane $C_{\omega X}(\tau)=C_{\omega Y}(\tau)$.}
    \label{fig:2}
\end{figure}

In Fig.~\ref{fig:2} we show the proper time evolution $C_{\omega X}$, $C_{\omega Y}$, and $C_{\omega Z}$. We present the results for two choices of transport coefficients, i.e., $\eta/s_0=1/4\pi$, $\chi_s=3/4\pi$, and $\eta/s_0=5/4\pi$, $\chi_s=10/4\pi$. Note that the evolution of $C_{\omega X}$, $C_{\omega Y}$, and $C_{\omega Z}$ does not directly depend on $\eta/s_0$ (see Eqs.~\eqref{equ63ver2}-\eqref{equ65ver2}), but in the spin hydrodynamic framework the evolution of these components of the spin chemical potential crucially depend on the proper time evolution of temperature ($T(\tau)$). Since the evolution of temperature depends on $\eta/s_0$, the proper time evolution of different components of the spin chemical potential is also affected by $\eta/s_0$. From this figure, we observe that $C_{\omega X}$, $C_{\omega Y}$, and $C_{\omega Z}$ decrease with proper time, and the decrease of these components is larger for higher values of different transport coefficients. Moreover, for the boost invariant system, the evolution equation of $C_{\omega X}$, and $C_{\omega Y}$ are identical (see Eqs.~\eqref{equ63ver2}-\eqref{equ64ver2}), i.e., $C_{\omega X}(\tau)=C_{\omega Y}(\tau)$. The equation governing $C_{\omega Z}$ does not contain any dissipative contributions from the spin transport coefficients, leading to a much slower decay compared to the transverse components. For different values of parameters that we consider here, we observe that (not shown here explicitly) if we start with unpolarized system, i.e., the initial values of $C_{\omega X}(\tau_0)=0$, $C_{\omega Y}(\tau_0)=0$, and $C_{\omega Z}(\tau_0)=0$, then after subsequent evolution the system remains unpolarized, i.e., $C_{\omega X}(\tau)=0$, $C_{\omega Y}(\tau)=0$, and $C_{\omega Z}(\tau)=0$, for all $\tau\geq\tau_0$.

\begin{figure}[]
    \centering
    \includegraphics[scale=0.55]{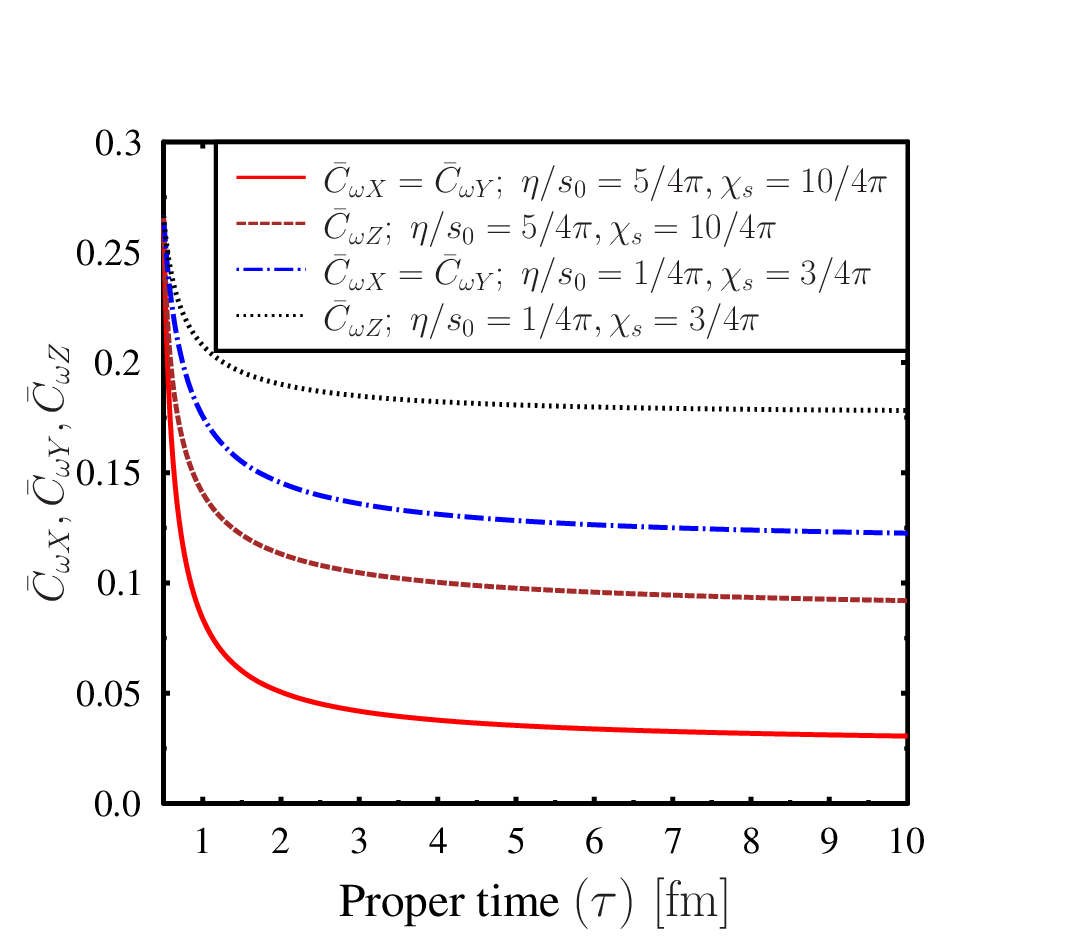}
    \caption{We show the the proper time evolution of dimensionless ratio $\bar{C}_{\omega X}\equiv C_{\omega X}/T$, $\bar{C}_{\omega Y} \equiv C_{\omega Y}/T$, and $\bar{C}_{\omega Z} \equiv C_{\omega Z}/T$. We observe that $\bar{C}_{\omega X}$, $\bar{C}_{\omega Y}$, and $\bar{C}_{\omega Z}$ remains below 1 for the entire evolution. Numerical solutions remain safely within the regime of small polarization limit. }
    \label{fignew}
\end{figure}

In the present calculation we consider the spin chemical potential $\omega^{\mu\nu}\sim \mathcal{O}(1)$, and also the temperature $T\sim\mathcal{O}(1)$. But we consider small polarization limit, i.e., $\omega^{\mu\nu}/T<1$ throughout the hydrodynamic evolution. This is to ensure that thermodynamic quantities are only quadratic in the spin chemical potential (Eq.~\eqref{equ24ver2}). To show the self-consistency of this truncation, we present the proper time evolution of dimensionless quantities, $\bar{C}_{\omega X}\equiv C_{\omega X}/T$, $\bar{C}_{\omega Y}\equiv C_{\omega Y}/T$, $\bar{C}_{\omega Z}\equiv C_{\omega Z}/T$ in Fig.~\ref{fignew}. In this figure, we consider all the parameter values that we took in Fig.~\ref{fig:2} , i.e., we choose the thermalization time scale $\tau_0 = 0.5$ fm to start the hydrodynamic evolution. At the initial time $\tau_0$ the initial temperature is $T(\tau_0)=T_0=300$ MeV and 
$C_{\omega X}(\tau_0)=C_{\omega Y}(\tau_0)=C_{\omega Z}(\tau_0)=80$ MeV. Therefore at $\tau=\tau_0$, the dimensionless variables $\bar{C}_{\omega X}(\tau_0)=\bar{C}_{\omega Y}(\tau_0)=\bar{C}_{\omega Z}(\tau_0)\sim 0.267<1$. From this figure, we observe that $\bar{C}_{\omega X}$, $\bar{C}_{\omega Y}$, and $\bar{C}_{\omega Z}$ remains less than one for the entire evolution of the system, validating our approximation during the hydrodynamic evolution. 

\section{Rate of Thermal Dilepton Production}
\label{SecVI}
We now apply the temperature profile ($T(\tau)$) to compute the thermal dilepton production from the medium. Dileptons are considered to be an important probe of the medium produced in heavy-ion collisions~\cite{Ryblewski:2015hea,Vujanovic:2013jpa}. Leptons interact only via electromagnetic interactions; hence, they have a low interaction cross-section and a longer mean free path. The dominant channel for dilepton production is quark-antiquark annihilation, $q\bar{q} \rightarrow \gamma^* \rightarrow \ell^+ \ell^-$ (similar to Drell-Yan process). For massless quarks rate of dilepton production is given by~\cite{Bhatt:2011kx,Singh:2018bih,Vogt:2007zz} 
\begin{align}
    & \frac{dN}{d^4x} = M^2
\int \frac{d^3p_1}{(2\pi)^3}
\frac{d^3p_2}{(2\pi)^3}
\frac{f(E_1) f(E_2)}{2E_1 E_2}
\, \sigma(M) 
\end{align}
where $\vec{p}_1$, $\vec{p}_2$ and $E_1 , E_2$ are the momenta and energy of the dileptons, respectively. $M$ is the invariant mass of the dilepton pair, and $\sigma(M)$ is the cross section of thermal dileptons. In the Born approximation for $N_f=2$, and $N_c=3$ the cross-section is given as, $\sigma(M)=\frac{80 \pi \alpha^2}{9M^2}$~\cite{Singh:2018bih}. $N_f$ is the light quark flavour, and $N_c$ is the color degree of freedom. $\alpha=1/137$. $f(E)$ is the Fermi-Dirac (FD) distribution function. In the limit $M \gg T$, one can replace the FD distribution function with Maxwell Boltzmann distribution function. In this limit, it can be shown that~\cite{Bhatt:2011kx,Singh:2018bih,Vogt:2007zz}  
\begin{align}
 & E \frac{dN}{d^4x \, d^3p \, dM^2} = \frac{1}{4}
\frac{M^2 \sigma(M)}{(2\pi)^5}\exp\left(-\frac{E}{T}\right)
\end{align}
Here, $E=E_1+E_2$ is the energy of $\ell^+$, $\ell^-$ pair. The effect of spin dynamics comes through the proper time evolution of temperature. The above equation is valid in the fluid rest frame, and this expression can be generalized for a general fluid frame, by replacing $E=p^0$, by $u_{\mu}p^{\mu}$, here
\begin{align}
p^{\mu}=\left(M_T\cosh{y},p_T\cos\phi,p_T\sin\phi, M_T\sinh{y}  \right)
\end{align}
$y$ is the particle rapidity, $\phi$ is the azimuthal angle, and $M_T=\sqrt{p_T^2+M^2}$. For a boost invariant system with $u^{\mu}\equiv (\cosh\eta_s,0,0,\sinh\eta_s)$, 
\begin{align}
E=u_{\mu}p^{\mu}=M_T\cosh(y-\eta_s).
\end{align}
For a boost invariant system, it can be shown that~\cite{Vogt:2007zz}, 
$d^4 x= \pi R^2\tau d\tau d\eta_s$, where $R$ is the transverse size, usually considered as the nuclear radius, and $R = 1.2\,A^{1/3} \mathrm{fm}$. We consider Gold Nuclei, with $A=197$. Moreover, 
$d^3p/E=2\pi p_T dp_T dy$, the differential dilepton production rates can be expressed as, 
\begin{align}
    & \frac{dN}{dM dy}= 4 M \pi^2 R^2\int_{\tau_0}^{\tau_{max}} \tau\, d\tau \int_{-\eta_{\min}}^{\eta_{\max}} d\eta_s \int p_T\, dp_T \nonumber\\
    & ~~~~~~~~~~~~~~~~~~~~~~~~~~~~\times\left( E \frac{dN}{d^{4}x\, d^{3}p\, dM^{2}} \right)
\end{align}
and
\begin{align}
    & \frac{dN}{p_T dp_TdM dy}= 4 M \pi^2 R^2\int_{\tau_0}^{\tau_{max}} \tau\, d\tau \int_{-\eta_{\min}}^{\eta_{\max}} d\eta_s \nonumber\\
    & ~~~~~~~~~~~~~~~~~~~~~~~~~~~~\times\left( E \frac{dN}{d^{4}x\, d^{3}p\, dM^{2}} \right)
\end{align}
For a boost invariant system the integrand of the above equations are independent of the azimuthal angle ($\phi$). Hence we have already performed the integration over the azimuthal angle ($\phi$) to obtain the above equations.
The above expressions are evaluated numerically to obtain the corresponding dilepton spectra. The initial proper time is set to $\tau_0 = 0.5\,\mathrm{fm}$. The upper limit of the proper time integration is denoted as $\tau_{\max}$.
$\tau_{\max}$ defines the proper time, when the temperature of the system 
reaches the quark-hadron transition temperature $T_c=150$ MeV~\cite{Singh:2018bih}, i.e., $T(\tau_{max})=T_c$. The space-time 
rapidity integration is performed within the limits $\eta_s \in [-5.3, 5.3]$~\cite{Singh:2018bih}. 

\begin{figure}
    \includegraphics[scale=0.55]{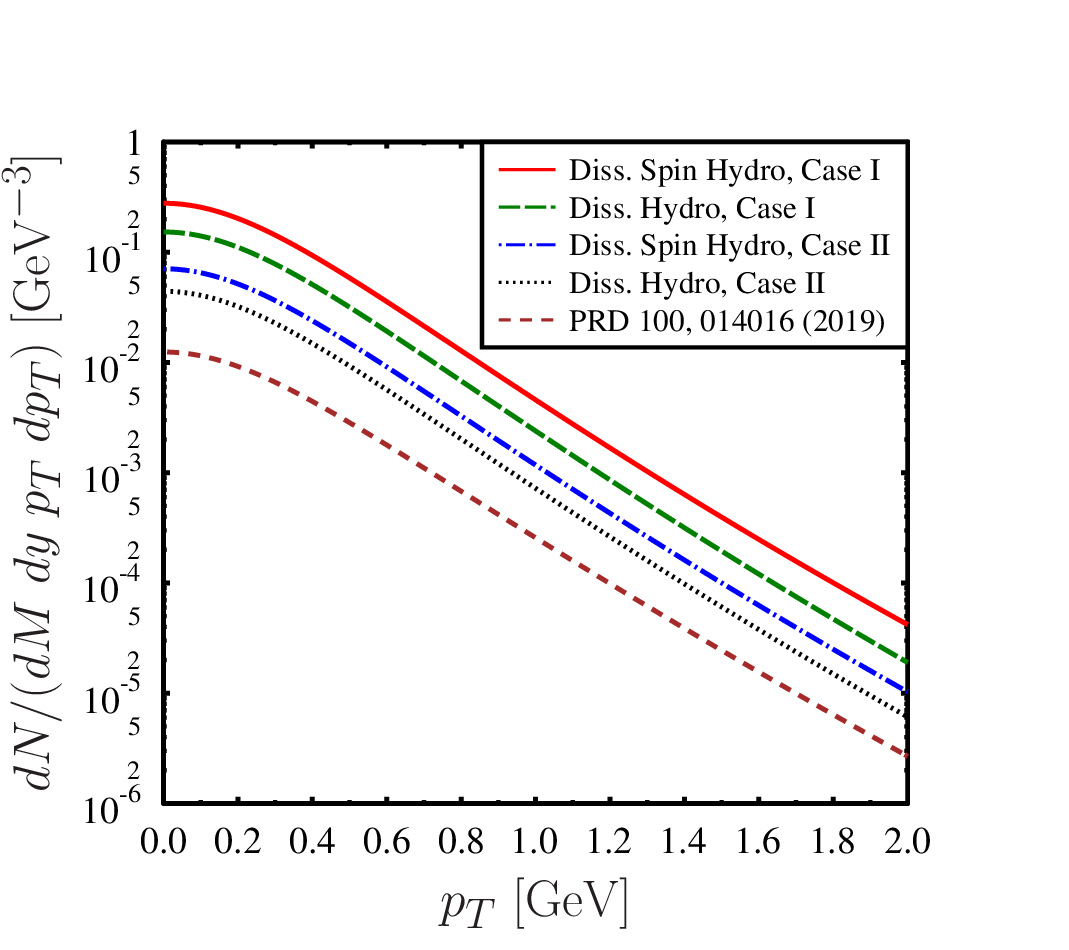}
    \caption{Dilepton production rate as a function of transverse momenta $p_T$. Here, the invariant mass is considered to be $M=0.4$ GeV. For the estimation of the dilepton production rate the temperature profile of the medium has been obtained for (a) Case I: dissipative spin hydrodynamics with $\eta/s_0=5/4\pi$, $\chi_s=10/4\pi$, dissipative hydrodynamics with $\eta/s_0=5/4\pi$, and  (b) Case II: dissipative spin hydrodynamics with $\eta/s_0=1/4\pi$, $\chi_s=3/4\pi$, dissipative hydrodynamics with $\eta/s_0=1/4\pi$. For the spin hydrodynamic case we consider $C_{\omega X}(\tau_0)=C_{\omega Y}(\tau_0)=C_{\omega Z}(\tau_0)=80$ MeV. We compare our results with the results obtained in Ref.~\cite{Singh:2018bih}. The brown dashed line represents the dilepton rate from a medium with non-vanishing vorticity $\omega_0=0.7$ fm$^{-1}$ obtained in Ref.~\cite{Singh:2018bih}.}
    \label{fig3}
\end{figure}

\begin{figure}
    \includegraphics[scale=0.55]{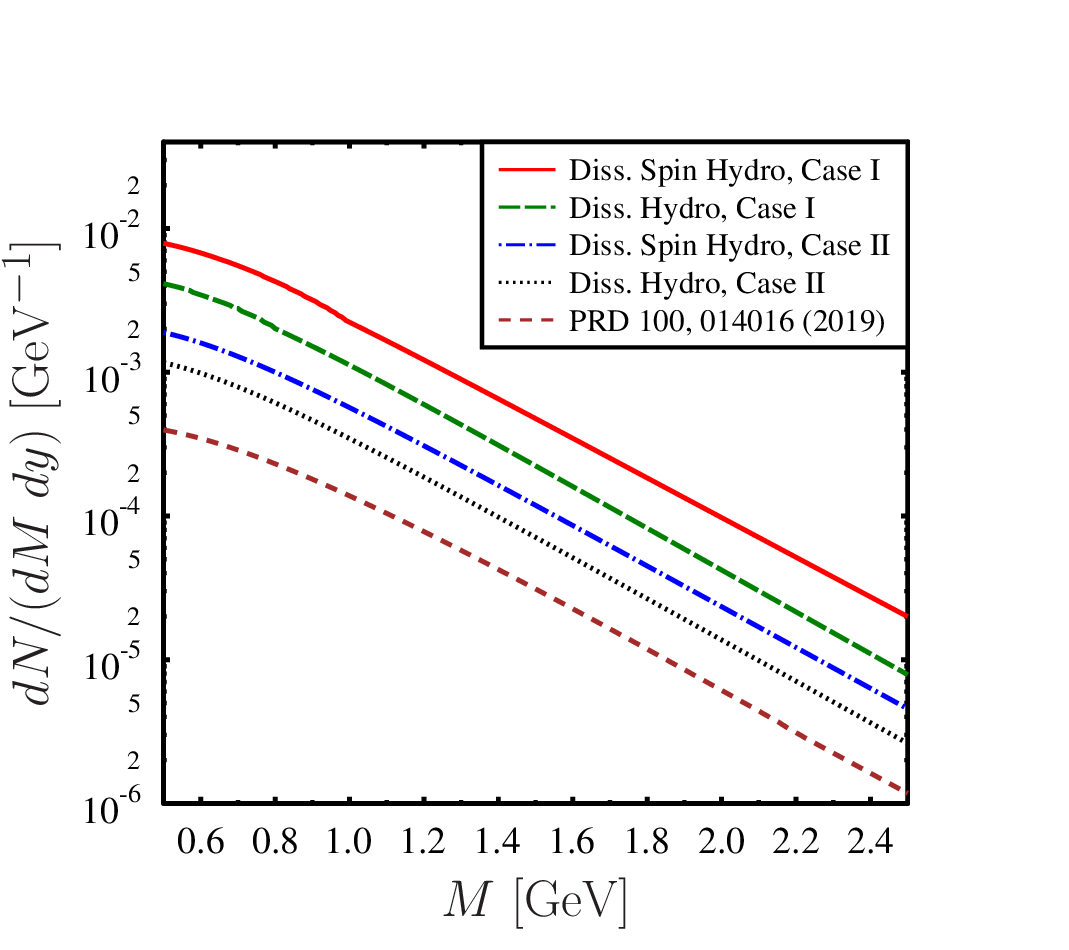}
    \caption{Dilepton production rate as a function of invariant mass $M$. Here, the transverse momentum has been considered in the range $0.5\leq p_T\leq 2 $ GeV. Here, four temperature profiles have also been used to obtain the rates. (a) Case I: dissipative spin hydrodynamics with $\eta/s_0=5/4\pi$, $\chi_s=10/4\pi$, dissipative hydrodynamics with $\eta/s_0=5/4\pi$, and (b) Case II: dissipative spin hydrodynamics with $\eta/s_0=1/4\pi$, $\chi_s=3/4\pi$, dissipative standard hydrodynamics with $\eta/s_0=1/4\pi$. For the spin hydrodynamic case we consider $C_{\omega X}(\tau_0)=C_{\omega Y}(\tau_0)=C_{\omega Z}(\tau_0)=80$ MeV. We compare our results with the results obtained in Ref.~\cite{Singh:2018bih} (brown dashed line) for a medium with non-vanishing vorticity $\omega_0=0.7$ fm$^{-1}$~\cite{Singh:2018bih}.}
    \label{fig4}
\end{figure}

In Figs.~\ref{fig3} and \ref{fig4}, we show the estimation of the dilepton rates. In Fig.~\ref{fig3}, we show the variation in the dilepton production rate $dN/(p_T dp_TdM dy)$ with transverse momentum $p_T$. For the estimation of $dN/(p_T dp_TdM dy)$, the invariant mass is considered to be $M=0.4$ GeV~\cite{Singh:2018bih}. In Fig.~\ref{fig4}, we show the variation in the dilepton production rate $dN/(dM dy)$ with invariant mass $M$. For the estimation of $dN/(dM dy)$, the transverse momentum has been considered in the range $0.5\leq p_T\leq 2 $ GeV~\cite{Singh:2018bih}. To obtain the results as shown in Figs.~\ref{fig3} and \ref{fig4} we have considered the particle rapidity $y=0$.

Note that medium temperature plays the crucial role in the estimation of dilepton rates. We use different temperature profiles $T(\tau)$, as shown in Fig.~\ref{fig:1}, for different scenarios. In Figs.~\ref{fig3} and \ref{fig4} the red solid line and green dashed line represent the scenario where $T(\tau)$ has been obtained from the dissipative spin hydrodynamic framework (Diss. Spin Hydro Case I), with $\eta/s_0=5/4\pi$, $\chi_s=10/4\pi$, and standard dissipative hydrodynamics (Diss. Hydro Case I) with $\eta/s_0=5/4\pi$, respectively. The blue dashed dotted line and black dotted line represent the scenario where $T(\tau)$ has been obtained from the dissipative spin hydrodynamic framework (Diss. Spin Hydro Case II), with $\eta/s_0=1/4\pi$, $\chi_s=3/4\pi$, and standard dissipative hydrodynamics (Diss. Hydro Case II) with $\eta/s_0=1/4\pi$, respectively. For the spin hydrodynamic case we consider the initial values of $C_{\omega X}(\tau_0)=C_{\omega Y}(\tau_0)=C_{\omega Z}(\tau_0)=80$ MeV. Here $\tau_0=0.5$ fm is the initial time of hydrodynamic evolution. From these figures, we observe that the lifetime of the partonic medium is very important for dilepton production. From Fig.~\ref{fig:1} we observe that for spin hydrodynamics, the decrease of temperature with proper time is slower compared to the standard dissipative hydrodynamics, hence the lifetime of the partonic medium is larger. With dissipative effects, this time scale is even larger. Hence, we observe an enhancement in the thermal dilepton production rate in the spin-hydrodynamic framework with stronger dissipative effects. We also compare our results with the results obtained in Ref.~\cite{Singh:2018bih}. In Ref.~\cite{Singh:2018bih}, the authors study the dilepton production in a partonic medium with finite vorticity. In Ref.~\cite{Singh:2018bih}, the authors also considered a boost-invariant system. In Figs.~\ref{fig3} and \ref{fig4}, brown dashed lines represent the dilepton rates from a medium with non-vanishing vorticity $\omega_0=0.7$ fm$^{-1}$ as obtained in Ref.~\cite{Singh:2018bih}. In 
Ref.~\cite{Singh:2018bih} authors argue that with increasing vorticity ($\omega_0$), the dilepton production rates decrease, which is opposite to our results. In our analysis, in the presence of spin chemical potential, the dilepton rate increases. This difference might arise from the theoretical frameworks; e.g., in the present calculation, we implement a spin hydrodynamic approach, in which temperature and spin evolution are coupled. Such a dynamical framework has not been considered in Ref.~\cite{Singh:2018bih}, the authors only considered the effect of vorticity in the thermodynamic relation, but not in the hydrodynamic evolution equation.     

\begin{figure}
    \centering
    \includegraphics[scale=0.55]{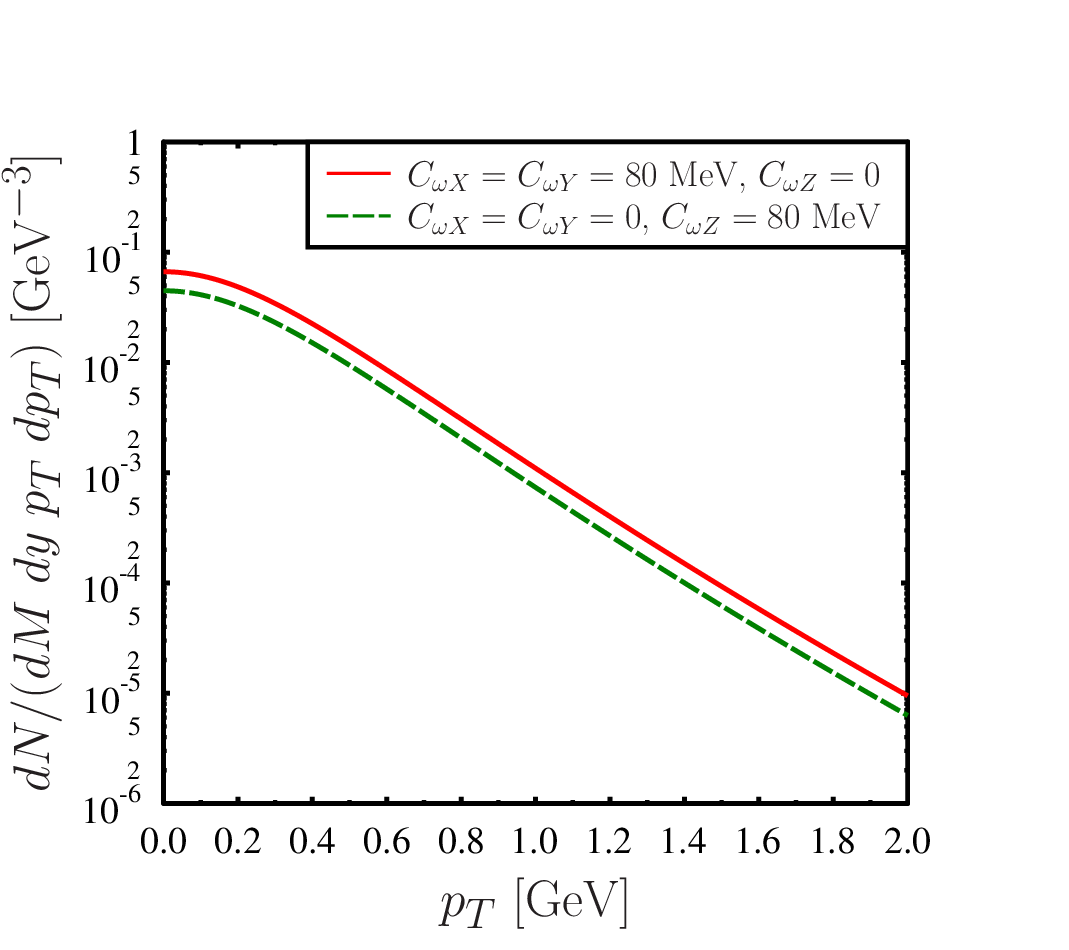}
    \caption{Dilepton production rate as a function of transverse momenta $p_T$. Here, the invariant mass is considered to be $M=0.4$ GeV. For this plot, $T(\tau)$ has been obtained by solving spin hydrodynamic equations with  $\eta/s_0=1/4\pi$, $\chi_s=3/4\pi$. The red solid line corresponds to the case where $C_{\omega X}(\tau_0)=C_{\omega Y} (\tau_0)=80$ MeV, $C_{\omega Z}(\tau_0)=0$. The green dashed line corresponds to $C_{\omega Z}(\tau_0)=80$ MeV,  $C_{\omega X}(\tau_0)=C_{\omega Y} (\tau_0)=0$.}
    \label{fig5}
\end{figure}

\begin{figure}
    \centering
    \includegraphics[scale=0.55]{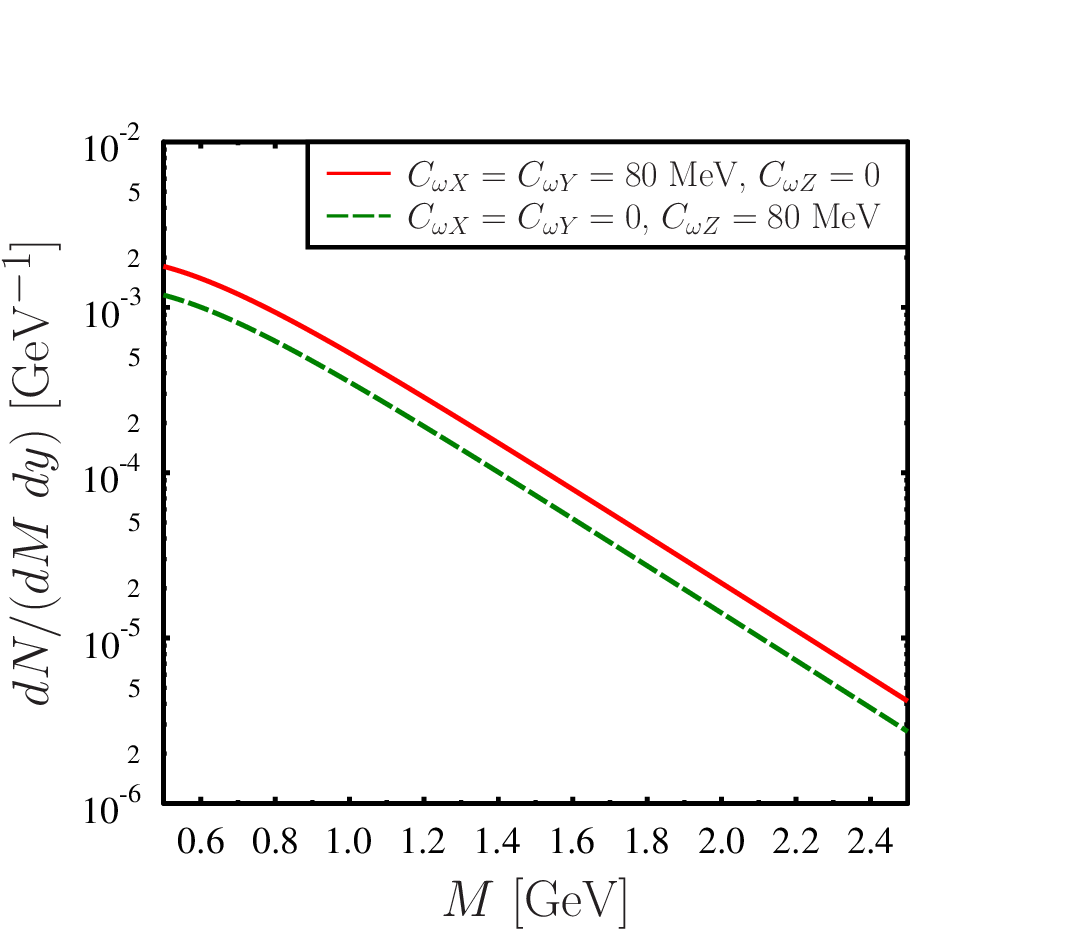}
    \caption{Dilepton production rate as a function of invariant mass $M$. Here, the transverse momentum has been considered in the range $0.5\leq p_T\leq 2 $ GeV. For this plot, $T(\tau)$ has been obtained by solving spin hydrodynamic equations with  $\eta/s_0=1/4\pi$, $\chi_s=3/4\pi$. The red solid line corresponds to the case where $C_{\omega X}(\tau_0)=C_{\omega Y} (\tau_0)=80$ MeV, $C_{\omega Z}=0$. The green dashed line corresponds to $C_{\omega Z}(\tau_0)=80$ MeV,  $C_{\omega X}(\tau_0)=C_{\omega Y} (\tau_0)=0$.}
    \label{fig6}
\end{figure}

On the experimental side different experimental
collaborations, e.g., the STAR RHIC BES data, ALICE data, give the total yield of the dileptons~\cite{Bailhache:2025kwa,Meninno:2020mjd,STAR:2015tnn}. In the QGP phase dileptons can originate from different processes, e.g., $q+\bar{q}\rightarrow \ell^+ + \ell^-$, $q+\bar{q}\rightarrow g+\ell^+ + \ell^-$, $g+q\rightarrow q +\ell^+ + \ell^-$. Moreover dileptons can also originate from the hadronic states, e.g., $\pi^0\rightarrow \gamma +e^{-}+e^{+}$, $\eta\rightarrow \gamma +e^{-}+e^{+}$, $\eta^{\prime}\rightarrow \gamma +e^{-}+e^{+}$, $\rho\rightarrow e^{-}+e^{+}$, $J/\psi\rightarrow e^{-}+e^{+}$, etc. Hence, one would need to incorporate different channels of dilepton production to compare the experimental data~\cite{Song:2018xca,Song:2018dvf,Song:2018wvd,Bailhache:2025kwa,Rapp:2013nxa,Rapp:1999ej}. In the present calculation, we estimate the effect of spin dynamics on the thermal dileptons originated from the process $q+\bar{q}\rightarrow \ell^+ +\ell^-$, which is only one of the process that contributes to the total dilepton yield. Moreover, chiral transition and thermal properties of hadronic medium can also be affected due to the medium vorticity~\cite{Wang:2018sur,Zhu:2025pxh,Farias:2025vss,Pradhan:2025pol,Mukherjee:2023ijv}, which may affect the dilepton production. Also, within the hydrodynamic simulation, the uncertainty in the initial condition and time evolution of various transport coefficients can affect the theoretical estimates. Therefore, for a proper comparison with the experimental data, we also need to look into the other processes that require separate detailed calculations.
In the absence of such an extensive calculation, instead of directly comparing with experimental data, we compare our results with existing theoretical estimates, such as those reported in Ref.~\cite{Singh:2018bih}.

In Sec.~\ref{SecIII} we argued that for a boost invariant system the electric-like components vanish ($\kappa^{\mu}=0$), and only magnetic-like components $(\omega^{\mu})$ become important. Moreover, the spin evolution is certainly determined by the proper time evolution of $C_{\omega X}$, $C_{\omega Y}$, and $C_{\omega Z}$. So far we have considered that $C_{\omega X}$, $C_{\omega Y}$, and $C_{\omega Z}$ are non-vanishing and have the same value at the initial time ($\tau_0$). But one can in principle consider two different cases, one with $C_{\omega X}=C_{\omega Y}\neq 0$, but $C_{\omega Z}=0$, and $C_{\omega X}=C_{\omega Y}= 0$, but $C_{\omega Z}\neq 0$. These configurations allow us to examine the effects of different components of the spin chemical potential on the system's evolution and dilepton rates. In Fig.~\ref{fig5} we show the variation of dilepton rate  $dN/(p_T dp_TdM dy)$ with $p_T$. To estimate $dN/(p_T dp_TdM dy)$ we have taken $M=0.4$ GeV. In Fig.~\ref{fig6} we show the variation of dilepton rate  $dN/(dM dy)$ with $M$. To estimate $dN/(dM dy)$ we have taken $0.5\leq p_T\leq 2$ GeV. For the estimation of dilepton rates, we have obtained the proper time evolution of medium temperature by solving spin hydrodynamic equations with $\eta/s_0=1/4\pi$, and $\chi_s=3/4\pi$. The red solid line in both Fig.~\ref{fig5}, and Fig.~\ref{fig6} corresponds to the case where we have set $C_{\omega X}(\tau_0)=C_{\omega Y}(\tau_0)=80$ MeV, but $C_{\omega Z}(\tau_0)=0$. Similarly, the green dashed line in Fig.~\ref{fig5}, and Fig.~\ref{fig6} corresponds to the case where we have considered $C_{\omega X}(\tau_0)=C_{\omega Y}(\tau_0)=0$, but $C_{\omega Z}(\tau_0)=80$ MeV. It can be shown that if any $C_{\omega i}(\tau_0)=0$, where $i\in \{X, Y, Z\}$, then for later times also, these coefficients remain zero. Although it has not been shown explicitly, it can be argued that for $C_{\omega X}(\tau_0)=C_{\omega Y}(\tau_0)=80 ~\text{MeV}, C_{\omega Z}(\tau_0)=0$ the medium temperature decreases slowly as compared to the case where $C_{\omega X}(\tau_0)=C_{\omega Y}(\tau_0)=0, C_{\omega Z}(\tau_0)=80$ MeV. Hence, when we have only the longitudinal component of the spin chemical potential ($C_{\omega Z}\neq 0$), the lifetime of the partonic medium is shorter, giving rise to a lower dilepton production rate, which can be observed in Fig.~\ref{fig5} and Fig.~\ref{fig6}.

 Some comment regarding the spin hydrodynamic framework with asymmetric energy-momentum tensor is in order here. We again emphasize that, for theoretical consistency, if one considers an anti-symmetric component in the energy-momentum tensor, then spin chemical potential ($\omega^{\mu\nu}$) should be considered as $\mathcal{O}(\partial)$~\cite{Hattori:2019lfp,Daher:2022xon,Biswas:2023qsw,Biswas:2022bht}. On the other hand, if the energy-momentum tensor is symmetric, then spin chemical potential ($\omega^{\mu\nu}$) should be considered as $\mathcal{O}(1)$~\cite{Dey:2024cwo}. We are considering spin chemical potential ($\omega^{\mu\nu}$) as $\mathcal{O}(1)$, hence we consider symmetric energy-momentum tensor or vice versa. The present calculation can not be generalized to incorporate an antisymmetric component in the energy-momentum tensor. There is a crucial distinction between spin hydrodynamic frameworks in which the spin chemical potential is considered as $\omega^{\mu\nu}\sim\mathcal{O}(1)$ and those in which it is considered as $\omega^{\mu\nu}\sim\mathcal{O}(\partial)$. In the former case, spin dissipation already appears in the Navier–Stokes limit, namely at first order in the gradient expansion. In contrast, in the framework with $\omega^{\mu\nu}\sim\mathcal{O}(\partial)$, spin dissipation does not contribute at the Navier–Stokes level~\cite{Hattori:2019lfp,Daher:2022xon,Biswas:2023qsw,Biswas:2022bht}. Therefore, in the spin hydrodynamic framework with $\omega^{\mu\nu}\sim\mathcal{O}(\partial)$, i.e., with an anti- symmetric component in the energy-momentum tensor, one would not observe the effects of spin dissipation/ spin transport coefficients on the medium evolution, and consequently on the dilepton production rates, at Navier-Stokes limit. In such a framework, the influence of spin transport coefficients on the system dynamics is expected to appear only at the second-order level or higher.

\section{Summary and outlook}
\label{SecVII}

In this work, we studied the boost-invariant evolution of a recently developed first-order spin hydrodynamic framework in which the spin chemical potential is treated as a leading-order hydrodynamic variable. Unlike formulations where the spin chemical potential is counted as a first-order quantity (order $\mathcal{O}(\partial)$) in the gradient expansion, the present framework allows one to consistently construct dissipative corrections to the spin tensor and to introduce spin transport coefficients at the Navier--Stokes level. We considered a baryon-free system with a symmetric energy-momentum tensor, so that the spin tensor is separately conserved. Adopting Bjorken flow, we derived the coupled evolution equations for the medium temperature and for the independent components of the spin chemical potential. Owing to boost invariance and the requirement that the total center-of-mass motion vanishes, the electric-like components of the spin chemical potential are identically zero, while only the magnetic-like components survive. The longitudinally expanding system is therefore characterized by three independent spin variables, denoted by $C_{\omega X}$, $C_{\omega Y}$, and $C_{\omega Z}$.

The dissipative spin hydrodynamic equations reveal an important qualitative difference between the transverse and longitudinal components of the spin chemical potential. The transverse components, $C_{\omega X}$ and $C_{\omega Y}$, receive dissipative contributions from the spin transport coefficients $\chi_2$ and $\chi_3$, and therefore decrease rapidly with proper time. In contrast, the longitudinal component $C_{\omega Z}$ does not receive such dissipative contributions and consequently decays much more slowly. We further observed that the decay of all spin components is indirectly affected by the shear viscosity through the temperature evolution of the medium. A central result of this work is that the spin degrees of freedom modify the temperature evolution of the longitudinally expanding medium. Compared to standard dissipative hydrodynamics, the presence of a non-vanishing spin chemical potential leads to a slower cooling of the system. For sufficiently large values of the transport coefficients, we even find a brief initial increase in the temperature before the subsequent cooling sets in, which is reminiscent of the reheating effect in the conventional Navier-Stokes hydrodynamics at early times.

Subsequently, we used the modified temperature profiles to estimate thermal dilepton production from quark-antiquark annihilation. Since dileptons are emitted throughout the evolution of the medium and interact only electromagnetically, they are particularly sensitive to the temperature history of the plasma. We found that the slower cooling in spin hydrodynamics enhances the dilepton production rate compared to the standard dissipative hydrodynamic case. This enhancement is visible both in the transverse-momentum spectrum and in the invariant-mass spectrum of the dileptons, and becomes more pronounced for larger values of the spin transport coefficients. The present study provides one of the first demonstrations of how dissipative spin dynamics can leave an imprint on thermal dilepton spectra. Although we employed a simplified equation of state and phenomenological choices for the spin transport coefficients, our results indicate that thermal dileptons may serve as an indirect probe of spin transport in the quark-gluon plasma.

We would like to emphasize that the present study is restricted to an idealized boost-invariant expansion (only longitudinal expansion) and is not intended as a quantitative description of the complete space-time evolution of the QGP. Consequently, the longer-lived temperature profile shown in Fig.~\ref{fig:1} should be interpreted as a demonstration that dynamical spin degrees of freedom can modify the cooling history, rather than as a direct prediction of the lifetime of the fireball produced in heavy-ion collisions. This is because of the absence of transverse expansion in the present calculation. 

Regarding compatibility with other observables, such as high-$p_T$ suppression, we note that these observables are sensitive to the full three-dimensional hydrodynamic evolution and are usually constrained through global fits involving realistic initial conditions, transverse expansion, hadronic freeze-out, and temperature-dependent transport coefficients. Any quantitative implementation of spin hydrodynamics in a realistic heavy-ion simulation would indeed require a reanalysis of the initial conditions and medium parameters to maintain agreement with existing hadronic and jet-quenching observables. In the standard model of heavy-ion collision, it is the appropriate initial condition, followed by a dissipative hydrodynamic modeling, and then the freezeout scenario, that together provide the complete evolution of the medium produced. The complete model predictions were then confronted with different observables. In the present calculation, we only discuss the effect of spin in the hydrodynamic stage, without investigating the effect of spin/non-zero angular momentum at the initial state, or at the freezeout stage. Note that the initial conditions that allow for non-vanishing angular momentum would be suitable to study the spin hydrodynamic framework in order to explain the spin observables. Not all initial state models that are used for spin-less hydrodynamics can account for non-vanishing angular momentum, e.g., the optical
Glauber initial condition does not provide non-vanishing orbital angular momentum at the initial stage. This is because in the optical Glauber model, at midrapidity, the initial energy density distribution is boost-invariant and symmetric about the y-axis. On the other hand, the AMPT model, which incorporates HIJING strings, can account for non-vanishing orbital angular momentum at the initial stage~\cite{Wu:2019eyi}. Hence, by naively tuning any initial condition model, one may not be able to reach to a correct conclusion. In a similar manner, the effect of vorticity should be incorporated at the freezeout stage. For any quantitative estimates of spin observables, one would need to implement the spin hydrodynamic framework with appropriate initial conditions and the freezeout prescription. However, any correct spin hydrodynamic model description should also be consistent with other non-spin observables, e.g., high $p_T$ suppression of patrons.

Because we are only dealing with the hydrodynamic part of the model, quantitatively it would be difficult to argue about the degree of tuning of the initial condition to avoid spoiling existing fits to hadronic data. On the other hand, it is also worth noting that the spin contribution in our calculations is controlled by the magnitude of the spin chemical potential. For the parameter choices considered here, the resulting temperature modification is moderate and primarily serves to illustrate the qualitative sensitivity of the evolution to spin dynamics. A quantitative assessment of the impact on hadronic spectra, flow observables, or jet quenching requires a full (3+1)-dimensional spin-hydrodynamic simulation and lies beyond the scope of the present work.


Looking forward, there are several possible directions for future work. A more realistic description would require the use of an equation of state obtained from lattice QCD or a microscopic quasiparticle model, together with microscopic estimates of the spin transport coefficients. It would also be interesting to include finite baryon density, transverse expansion, and realistic initial conditions relevant for heavy-ion collisions. In addition, the spin hydrodynamic framework could be extended to investigate the effect of spin evolution on other observables such as thermal photon production. Such studies may provide further insight into the role of spin dynamics in relativistic heavy-ion collisions.


\section*{Acknowledgments}

S.D. and A.D. acknowledge the New Faculty Seed Grant (NFSG), NFSG/PIL/2024/P3825, provided by the Birla Institute of Technology and Science Pilani, Pilani Campus, India. A.D. and A.J. acknowledge the Anusandhan National Research Foundation (ANRF), Advanced Research Grant (ARG), project number: ANRF/ARG/2025/000691/PS.  A.J. gratefully acknowledges the Department of Atomic Energy (DAE), India, for financial support. 

\appendix

\section{Conservation of energy-momentum tensor for a boost invariant system}
\label{appendix1}
For Bjorken flow in the Cartesian coordinates, $u^{\mu} = (\cosh \eta_s, 0, 0, \sinh \eta_s)$.
The Cartesian coordinate $(t,x,y,z)$ can be transformed in to the Milne coordinate  $(\tau,x,y,\eta_s)$
using the following transformations $t=\tau \cosh \eta_s$, $z=\tau \sinh \eta_s$. Here $\eta_s =(1/2)\ln[(t+z)/(t-z)]$ is the spacetime rapidity, and the proper time $\tau=\sqrt{t^2-z^2}$. The derivative with respect to time ($t$) and $z$ coordinate can be expressed as, 
\begin{align}
& \partial_t=\partial_0=\cosh\eta_s\partial_{\tau}-\frac{\sinh\eta_s}{\tau}\partial_{\eta_s}\label{equA1new}\\
& \partial_z=\partial_3=-\sinh\eta_s\partial_{\tau}+\frac{\cosh\eta_s}{\tau}\partial_{\eta_s} \label{equA2new}
\end{align}
Using the explicit expression of $u^{\mu}$, and Eqs.~\eqref{equA1new}-\eqref{equA2new}, it can be shown that, 
\begin{align}
& u^{\mu}\partial_{\mu}f(\tau,\eta_s)=\frac{\partial f(\tau,\eta_s)}{\partial\tau}\label{equA3new}\\ 
& \theta =\partial_{\mu}u^{\mu}=\nabla_{\mu}u^{\mu}=\frac{1}{\tau} \label{equA4new}\\
& \nabla^{0}=-\frac{\sinh\eta_s}{\tau}\frac{\partial}{\partial\eta_s} \label{equA5new}\\
& \nabla^{3}=-\frac{\cosh\eta_s}{\tau}\frac{\partial}{\partial\eta_s} \label{equA6new}
\end{align}
Using the expression of $\pi^{\mu\nu}$, 
\begin{align}
   & \pi^{\mu \nu}=2 \eta\sigma^{\mu\nu}=2\eta\bigg[\frac{1}{2}(\nabla^\mu u^\nu+\nabla^{\nu}u^\mu)-\frac{1}{3}\Delta^{\mu \nu}\theta \bigg]\label{equA7new}
\end{align}
it can be shown that, 
\begin{align}
&  \pi^{\mu\nu}\partial_{\mu}u_{\nu}=  \pi^{\mu\nu}\nabla_{\mu}u_{\nu}\nonumber\\
= & \eta\bigg[(\nabla^{\mu}u^{\nu})\nabla_{\mu}u_{\nu}+(\nabla^{\nu}u^{\mu})\nabla_{\nu}u_{\mu}-\frac{2}{3}\theta \nabla^{\nu}u_{\nu}\bigg]\nonumber\\
= & \eta\bigg[\frac{1}{\tau^2}+\frac{1}{\tau^2}-\frac{2}{3}\frac{1}{\tau^2}\bigg]=\frac{4}{3}\frac{\eta}{\tau^2}.\label{equA8new}
\end{align}
also, 
\begin{align}
\Pi=\zeta\theta=\frac{\zeta}{\tau}
\label{equA9new2}
\end{align}
Using Eqs.~\eqref{equA3new}, \eqref{equA4new}, \eqref{equA8new}, and \eqref{equA9new2} in Eq.~\eqref{equ28ver2} we find,  
\begin{align}
       \frac{d\varepsilon}{d\tau}+\frac{\varepsilon+P}{\tau}-\bigg(\frac{4}{3}\frac{\eta}{\tau^2}+\frac{\zeta}{\tau^2}\bigg)=0.
    \label{equA9new}
\end{align}
Moreover for the boost invariant system,
\begin{align}
& \left(u^{\mu} \partial_{\mu}\right) u^{\alpha}=0, \\
& \Delta^{\alpha\mu}\partial_{\mu} \left(P-\Pi\right)=0. 
\end{align}
Note that $u^{\mu}$ only depend on space-time rapidity $\eta_s$. On the other hand $P$, and $\Pi$ only depends on the proper time ($\tau$). For a boost invariant system different components of $\pi^{\mu\nu}$ are, 
\begin{align}
& \pi^{01}=0; \pi^{02}=0\\
& \pi^{00}=-\frac{4}{3}\frac{\eta}{\tau}(\sinh\eta_s)^2; \pi^{03}=-\frac{4}{3}\frac{\eta}{\tau}\sinh\eta_s \cosh\eta_s\\
& \pi^{11}=\frac{2}{3}\frac{\eta}{\tau}; \pi^{12} =0; \pi^{13} =0\\
& \pi^{22}=\frac{2}{3}\frac{\eta}{\tau}; \pi^{23} =0; \pi^{33}=-\frac{4}{3}\frac{\eta}{\tau}(\cosh\eta_s)^2
\end{align}
Using the above expression of different components of $\pi^{\mu\nu}$ one can show that, for a boost invariant system,
\begin{align}
\Delta^{\alpha}_{~\nu}\partial_{\mu}\pi^{\mu\nu}=0.
\end{align}
Therefore for a boost invariant system Eq.~\eqref{equ29ver2} is trivially satisfied.

\section{Derivation of Eqs.~\eqref{equ63ver2}-\eqref{equ65ver2}}
\label{appendix2}
Contracting Eq.~\eqref{equ54ver2} with $u_{\alpha}X_{\beta}$ one finds,
\begin{align}
   u_{\alpha}X_{\beta}\bigg[ \frac{\partial S^{\alpha \beta}}{\partial \tau} +\frac{S^{\alpha \beta}}{\tau}+ \partial_\mu S^{\mu\alpha\beta}_{(1)}\bigg]=0.
    \label{}
\end{align}
Now, 
\begin{align}
& u_{\alpha}X_{\beta}\frac{\partial S^{\alpha \beta}}{\partial \tau} = \frac{\partial}{\partial\tau}\bigg[u_{\alpha}X_{\beta}S^{\alpha\beta}\bigg]= \frac{\partial}{\partial\tau}\bigg[S_0(T) u_{\alpha}X_{\beta}\omega^{\alpha\beta}\bigg]\nonumber\\
& = \frac{\partial}{\partial\tau}\bigg[S_0(T)u_{\alpha}X_{\beta}\epsilon^{\alpha\beta\gamma\delta}u_{\gamma}\omega_{\delta}\bigg]=0.
\end{align}
Similarly it can be shown that, $u_{\alpha}X_{\beta}S^{\alpha\beta}=0$. Moreover, 
\begin{align}
& u_{\alpha}X_{\beta}\partial_{\mu}S^{\mu\alpha\beta}_{(1)}= u_{\alpha}X_{\beta}\partial_{\mu}\bigg[u^{\alpha}\tau^{\mu\beta}_{(s)}-u^{\beta}\tau^{\mu\alpha}_{(s)}\bigg]\nonumber\\
& ~~~~~~~~~~~~~~~~~~~+u_{\alpha}X_{\beta}\partial_{\mu}\bigg[u^{\alpha}\tau^{\mu\beta}_{(a)}-u^{\beta}\tau^{\mu\alpha}_{(a)}\bigg]\nonumber\\
& = X_{\beta}\partial_{\mu}\tau^{\mu\beta}_{(s)}+X_{\beta}\partial_{\mu}\tau^{\mu\beta}_{(a)}\nonumber\\
& = -\partial_{0}\tau^{0 1}_{(s)}-\partial_{3}\tau^{31}_{(s)}-\partial_{0}\tau^{0 1}_{(a)}-\partial_{3}\tau^{31}_{(a)}=0. 
\end{align}
here we have used the expression of $\partial_{0}$, $\partial_{3}$, $\tau^{01}_{(s)}$, $\tau^{01}_{(a)}$, $\tau^{31}_{(s)}$, and $\tau^{31}_{(a)}$.
Similarly it can be shown that, 
\begin{align}
   u_{\alpha}Y_{\beta}\bigg[ \frac{\partial S^{\alpha \beta}}{\partial \tau} +\frac{S^{\alpha \beta}}{\tau}+ \partial_\mu S^{\mu\alpha\beta}_{(1)}\bigg]=0.
    \label{}
\end{align}
and, 
\begin{align}
   u_{\alpha}Z_{\beta}\bigg[ \frac{\partial S^{\alpha \beta}}{\partial \tau} +\frac{S^{\alpha \beta}}{\tau}+ \partial_\mu S^{\mu\alpha\beta}_{(1)}\bigg]=0.
    \label{}
\end{align}

Now contracting Eq.~\eqref{equ54ver2} with $X_{\alpha}Y_{\beta}$ one finds,
\begin{align}
   X_{\alpha}Y_{\beta}\bigg[ \frac{\partial S^{\alpha \beta}}{\partial \tau} +\frac{S^{\alpha \beta}}{\tau}+ \partial_\mu S^{\mu\alpha\beta}_{(1)}\bigg]=0.
    \label{equB6new}
\end{align}
Here, 
\begin{align}
& X_{\alpha}Y_{\beta}S^{\alpha\beta} = S_0(T)\epsilon^{\alpha\beta\gamma\delta}X_{\alpha}Y_{\beta}u_{\gamma}\omega_{\delta}\nonumber\\
& = S_0(T)\bigg[\epsilon^{1203}u_0\omega_3+\epsilon^{1230}u_3\omega_0\bigg]\nonumber\\
& = S_0(T) \bigg[-u^0\omega^3+u^3\omega^0\bigg]= -S_0(T)C_{\omega Z}.
\label{equB7new}
\end{align}
Moreover, 
\begin{align}
& X_{\alpha}Y_{\beta}\partial_{\mu}\bigg[u^{\alpha}\tau^{\mu\beta}_{(s)}-u^{\beta}\tau^{\mu\alpha}_{(s)}\bigg]=0,\label{equB8new}\\
& X_{\alpha}Y_{\beta}\partial_{\mu}\bigg[u^{\alpha}\tau^{\mu\beta}_{(a)}-u^{\beta}\tau^{\mu\alpha}_{(a)}\bigg]=0.\label{equB9new} 
\end{align}
Using Eqs.~\eqref{equB7new}-\eqref{equB9new} in Eq.~\eqref{equB6new} we find, 
\begin{align}
& \frac{\partial}{\partial\tau}\bigg[S_0(T)C_{\omega Z}\bigg]+\frac{S_0(T)C_{\omega Z}}{\tau}=0, \nonumber\\
\implies & \frac{d C_{\omega Z}}{d \tau} +C_{\omega Z} \bigg(\frac{S_0'(T)}{S_0(T)}\frac{dT}{d\tau}+\frac{1}{\tau}\bigg)=0.
\end{align}

Contracting Eq.~\eqref{equ54ver2} with $X_{\alpha}Z_{\beta}$ one finds,
\begin{align}
   X_{\alpha}Z _{\beta}\bigg[ \frac{\partial S^{\alpha \beta}}{\partial \tau} +\frac{S^{\alpha \beta}}{\tau}+ \partial_\mu S^{\mu\alpha\beta}_{(1)}\bigg]=0.
    \label{equB11new}
\end{align}
It can be shown that, 
\begin{align}
 & X_{\alpha}Z_{\beta}S^{\alpha\beta}=  X_{\alpha}Z_{\beta} S_0(T)\epsilon^{\alpha\beta\gamma\delta}u_{\gamma}\omega_{\delta}\nonumber\\
= & S_0(T) X^{1}Z^{3}u^{0}\omega^2- S_0(T) X^{1}Z^{0}u^{3}\omega^2\nonumber\\
= & S_0(T)C_{\omega Y}.
\label{equB12new}
\end{align}
Furthermore, 
\begin{align}
& X_{\alpha}Z_{\beta}\partial_{\mu}\bigg[u^{\alpha}\tau^{\mu\beta}_{(s)}-u^{\beta}\tau^{\mu\alpha}_{(s)}\bigg]= -X_{\alpha}\tau^{\mu\alpha}_{(s)}Z_{\beta}\partial_{\mu}u^{\beta}\nonumber\\
& = \tau^{\mu 1}_{(s)}Z_{\beta}\partial_{\mu}u^{\beta}= \tau^{\mu 1}_{(s)}\bigg[Z^0\partial_{\mu}u^0-Z^3\partial_{\mu}u^3\bigg]\nonumber\\
& = \tau^{01}_{(s)}\sinh\eta_s\partial_0(\cosh\eta_s)+\tau^{31}_{(s)}\sinh\eta_s\partial_3(\cosh\eta_s)\nonumber\\
& - \tau^{01}_{(s)}\cosh\eta_s\partial_0(\sinh\eta_s)-\tau^{31}_{(s)}\cosh\eta_s\partial_3(\sinh\eta_s)\nonumber\\
& = \frac{\chi_2\beta C_{\omega Y}}{\tau^2}.
\label{equB13new}
\end{align}
Similarly it can be shown that, 
\begin{align}
& X_{\alpha}Z_{\beta}\partial_{\mu}\bigg[u^{\alpha}\tau^{\mu\beta}_{(a)}-u^{\beta}\tau^{\mu\alpha}_{(a)}\bigg] = \tau^{\mu 1}_{(a)}Z_{\beta}\partial_{\mu}u^{\beta}\nonumber\\
& ~~~~~~~~~~~~~~~~~~~~~~~~~~~~~~~~~~~= \frac{\chi_3\beta C_{\omega Y}}{\tau^2}. 
\label{equB14new}
\end{align}
Using Eqs.~\eqref{equB12new}-\eqref{equB14new} in Eq.~\eqref{equB11new} we find the evolution equation for $C_{\omega Y}$,
\begin{align}
& \frac{d}{d\tau}\bigg[S_0(T)C_{\omega Y}\bigg]+\frac{S_0(T)C_{\omega Y}}{\tau}\nonumber\\
& ~~~~~~~~~~~~~~~~~~~~~~~~~+(\chi_2+\chi_3)\frac{\beta C_{\omega Y}}{\tau^2}=0. \nonumber\\
\implies & \frac{d C_{\omega Y}}{d \tau} +C_{\omega Y} \bigg(\frac{S_0'(T)}{S_0(T)}\frac{dT}{d\tau}+\frac{1}{\tau}\nonumber\\
& ~~~~~~~~~~~~~~~~~~~~~~~~~~~~~~~+ \frac{(\chi_2+\chi_3)}{\tau^2 T S_0(T)}\bigg)=0.
\end{align}
Similarly, contracting Eq.~\eqref{equ54ver2} with $Y_{\alpha}Z_{\beta}$,
\begin{align}
   Y_{\alpha}Z _{\beta}\bigg[ \frac{\partial S^{\alpha \beta}}{\partial \tau} +\frac{S^{\alpha \beta}}{\tau}+ \partial_\mu S^{\mu\alpha\beta}_{(1)}\bigg]=0,
    \label{}
\end{align}
it can be shown that,
\begin{align}
& \frac{d}{d\tau}\bigg[S_0(T)C_{\omega X}\bigg]+\frac{S_0(T)C_{\omega X}}{\tau}\nonumber\\
& ~~~~~~~~~~~~~~~~~~~~~~~~~+(\chi_2+\chi_3)\frac{\beta C_{\omega X}}{\tau^2}=0. \nonumber\\
\implies & \frac{d C_{\omega X}}{d \tau} +C_{\omega X} \bigg(\frac{S_0'(T)}{S_0(T)}\frac{dT}{d\tau}+\frac{1}{\tau}\nonumber\\
& ~~~~~~~~~~~~~~~~~~~~~~~~~~~~~~~+ \frac{(\chi_2+\chi_3)}{\tau^2 T S_0(T)}\bigg)=0.
\end{align}

\section{Derivation of Eq.~\eqref{equ66ver2}}
\label{appendix3}
Using the expression of $\varepsilon(T,\omega^{\mu\nu})$ (Eq.~\eqref{equ51ver2}) we find, 
\begin{align}
\frac{d\varepsilon}{d\tau}& =\frac{d\varepsilon_0}{dT}\frac{dT}{d\tau}+\bigg[2S_0^{\prime}(T)+T S_0^{\prime\prime}(T)\bigg]C^2\frac{dT}{d\tau}\nonumber\\
&~~~~~~~+ 2\bigg[S_0(T)+TS_0^{\prime}(T)\bigg]C\frac{dC}{d\tau} \nonumber\\
& = A \frac{dT}{d\tau}+B C \frac{dC}{d\tau}.
\label{equC1new}
\end{align}
Here, 
\begin{align}
    &A= \frac{d \varepsilon_0}{dT}+\bigg[2S_0'(T)+TS_0'' (T)\bigg] C^2,  \\ 
    &B=2\bigg[S_0(T)+TS_0'(T)\bigg].
\end{align}
Moreover $C^2=C_{\omega X}^2+C_{\omega Y}^2+C_{\omega Z}^2$, which implies, 
\begin{align}
C\frac{dC}{d\tau}= C_{\omega X}\frac{dC_{\omega X}}{d\tau}+C_{\omega Y}\frac{dC_{\omega Y}}{d\tau}+C_{\omega Z}\frac{dC_{\omega Z}}{d\tau}.
\label{equC4new}
\end{align}
Using Eqs.~\eqref{equ63ver2}-\eqref{equ65ver2} in Eq.~\eqref{equC4new} we find, 
\begin{align}
& C\frac{dC}{d\tau}=-C^2\bigg[\frac{S_0^{\prime}(T)}{S_0(T)}\frac{dT}{d\tau}+\frac{1}{\tau}\bigg]\nonumber\\
&~~~~~~~~~~~~~~~~~~-\left(C_{\omega X}^2+C_{\omega Y}^2\right)\frac{\chi_2+\chi_3}{S_0(T)}\frac{1}{T\tau^2}.
\label{equC5new}
\end{align}
Using Eq.~\eqref{equC5new} in Eq.~\eqref{equC1new} we get, 
\begin{align}
\frac{d\varepsilon}{d\tau}&=\bigg[A-B C^2\frac{S_0^{\prime}(T)}{S_0(T)}\bigg]\frac{dT}{d\tau}-\frac{BC^2}{\tau}\nonumber\\
& ~~~~~~~~~~~~-B \left(C_{\omega X}^2+C_{\omega Y}^2\right)\frac{\chi_2+\chi_3}{S_0(T)}\frac{1}{T\tau^2}.
\label{equC6new}
\end{align}
Using Eq.~\eqref{equ50ver2}-\eqref{equ51ver2} we find, 
\begin{align}
\frac{\varepsilon+P}{\tau}& =\frac{\varepsilon_0+P_0}{\tau}+\bigg[2S_0(T)+TS_0^{\prime}(T)\bigg]\frac{C^2}{\tau}\nonumber\\
& =\frac{Ts_0}{\tau}+\bigg[2S_0(T)+TS_0^{\prime}(T)\bigg]\frac{C^2}{\tau}.
\label{equC7new}
\end{align}
Finally using Eq.~\eqref{equC6new}, and Eq.~\eqref{equC7new} in Eq.~\eqref{equ53ver2} we find the proper time evolution of temperature, 
\begin{align}
& \bigg[A-BC^2\frac{S_0'(T)}{S_0(T)}\bigg] \frac{dT}{d\tau}-\frac{BC^2}{\tau}-\frac{B\left(C_{\omega X}^2+C_{\omega Y}^2\right)}{T\tau^2}\chi_s\nonumber\\
& ~~~~+\frac{Ts_0(T)}{\tau}+\bigg[2S_0(T)+TS_0^{\prime}(T)\bigg]\frac{C^2}{\tau}\nonumber\\
& ~~~~~~~~~~~~~~~~~~~~~~~~~~~~-\frac{s_0}{\tau^2}\bigg(\frac{4\eta}{3 s_0}+\frac{\zeta}{s_0}\bigg)=0.
\label{}
\end{align}
Here we define the dimensionless ratio $\chi_s=(\chi_2+\chi_3)/S_0(T)$.

\bibliography{ref.bib}{}

\providecommand{\href}[2]{#2}\begingroup\raggedright\begin{thebibliography}{10}

\bibitem{STAR:2017ckg}
{\bfseries STAR} Collaboration, L.~Adamczyk {\em et~al.}, ``{Global $\Lambda$ hyperon polarization in nuclear collisions: evidence for the most vortical fluid},'' \href{http://dx.doi.org/10.1038/nature23004}{{\em Nature} {\bfseries 548} (2017) 62--65},
\href{http://arxiv.org/abs/1701.06657}{{\ttfamily arXiv:1701.06657 [nucl-ex]}}.

\bibitem{STAR:2018gyt}
{\bfseries STAR} Collaboration, J.~Adam {\em et~al.}, ``{Global polarization of $\Lambda$ hyperons in Au+Au collisions at $\sqrt{s_{_{NN}}}$ = 200 GeV},'' \href{http://dx.doi.org/10.1103/PhysRevC.98.014910}{{\em Phys. Rev. C} {\bfseries 98} (2018) 014910}, \href{http://arxiv.org/abs/1805.04400}{{\ttfamily arXiv:1805.04400 [nucl-ex]}}.

\bibitem{STAR:2019erd}
{\bfseries STAR} Collaboration, J.~Adam {\em et~al.}, ``{Polarization of $\Lambda$ ($\bar{\Lambda}$) hyperons along the beam direction in Au+Au collisions at $\sqrt{s_{_{NN}}}$ = 200 GeV},'' \href{http://dx.doi.org/10.1103/PhysRevLett.123.132301}{{\em Phys. Rev. Lett.} {\bfseries 123} no.~13, (2019) 132301}, \href{http://arxiv.org/abs/1905.11917}{{\ttfamily arXiv:1905.11917 [nucl-ex]}}.

\bibitem{ALICE:2019onw}
{\bfseries ALICE} Collaboration, S.~Acharya {\em et~al.}, ``{Global polarization of $\Lambda \bar \Lambda$ hyperons in Pb-Pb collisions at $\sqrt {s_{NN}}$ = 2.76 and 5.02 TeV},'' \href{http://dx.doi.org/10.1103/PhysRevC.101.044611}{{\em Phys. Rev. C} {\bfseries 101} no.~4, (2020) 044611}, \href{http://arxiv.org/abs/1909.01281}{{\ttfamily arXiv:1909.01281 [nucl-ex]}}.

\bibitem{Kornas:2020qzi}
{\bfseries HADES} Collaboration, F.~J. Kornas, ``{$ \Lambda $ Polarization in Au+Au Collisions at $ \sqrt s_{NN} = 2.4\,{\text{GeV}} $ Measured with HADES},'' \href{http://dx.doi.org/10.1007/978-3-030-53448-6_68}{{\em Springer Proc. Phys.} {\bfseries 250} (2020) 435--439}.

\bibitem{Liang:2004ph}
Z.-T. Liang and X.-N. Wang, ``{Globally polarized quark-gluon plasma in non-central A+A collisions},'' \href{http://dx.doi.org/10.1103/PhysRevLett.94.102301, 10.1103/PhysRevLett.96.039901}{{\em Phys. Rev. Lett.} {\bfseries 94} (2005) 102301}, \href{http://arxiv.org/abs/nucl-th/0410079}{{\ttfamily arXiv:nucl-th/0410079 [nucl-th]}}.
[Erratum: Phys. Rev. Lett.96,039901(2006)].

\bibitem{Becattini:2007sr}
F.~Becattini, F.~Piccinini, and J.~Rizzo, ``{Angular momentum conservation in heavy ion collisions at very high energy},'' \href{http://dx.doi.org/10.1103/PhysRevC.77.024906}{{\em Phys. Rev.} {\bfseries C77} (2008) 024906},
\href{http://arxiv.org/abs/0711.1253}{{\ttfamily arXiv:0711.1253 [nucl-th]}}.

\bibitem{Gao:2007bc}
J.-H. Gao, S.-W. Chen, W.-T. Deng, Z.-T. Liang, Q.~Wang, and X.-N. Wang, ``{Global quark polarization in non-central A+A collisions},'' \href{http://dx.doi.org/10.1103/PhysRevC.77.044902}{{\em Phys. Rev.} {\bfseries C77} (2008) 044902},
\href{http://arxiv.org/abs/0710.2943}{{\ttfamily arXiv:0710.2943 [nucl-th]}}.

\bibitem{Huang:2011ru}
X.-G. Huang, P.~Huovinen, and X.-N. Wang, ``{Quark Polarization in a Viscous Quark-Gluon Plasma},'' \href{http://dx.doi.org/10.1103/PhysRevC.84.054910}{{\em Phys. Rev. C} {\bfseries 84} (2011) 054910}, \href{http://arxiv.org/abs/1108.5649}{{\ttfamily arXiv:1108.5649 [nucl-th]}}.

\bibitem{Becattini:2013fla}
F.~Becattini, V.~Chandra, L.~Del~Zanna, and E.~Grossi, ``{Relativistic distribution function for particles with spin at local thermodynamical equilibrium},'' \href{http://dx.doi.org/10.1016/j.aop.2013.07.004}{{\em Annals Phys.} {\bfseries 338} (2013) 32--49}, \href{http://arxiv.org/abs/1303.3431}{{\ttfamily arXiv:1303.3431 [nucl-th]}}.

\bibitem{Fang:2016vpj}
R.-H. Fang, L.-G. Pang, Q.~Wang, and X.-N. Wang, ``{Polarization of massive fermions in a vortical fluid},'' \href{http://dx.doi.org/10.1103/PhysRevC.94.024904}{{\em Phys. Rev.} {\bfseries C94} no.~2, (2016) 024904},
\href{http://arxiv.org/abs/1604.04036}{{\ttfamily arXiv:1604.04036 [nucl-th]}}.

\bibitem{Voloshin:2004ha}
S.~A. Voloshin, ``{Polarized secondary particles in unpolarized high energy hadron-hadron collisions?},''
\href{http://arxiv.org/abs/nucl-th/0410089}{{\ttfamily arXiv:nucl-th/0410089 [nucl-th]}}.

\bibitem{Betz:2007kg}
B.~Betz, M.~Gyulassy, and G.~Torrieri, ``{Polarization probes of vorticity in heavy ion collisions},'' \href{http://dx.doi.org/10.1103/PhysRevC.76.044901}{{\em Phys. Rev.} {\bfseries C76} (2007) 044901},
\href{http://arxiv.org/abs/0708.0035}{{\ttfamily arXiv:0708.0035 [nucl-th]}}.

\bibitem{Becattini:2015ska}
F.~Becattini, G.~Inghirami, V.~Rolando, A.~Beraudo, L.~Del~Zanna, A.~De~Pace, M.~Nardi, G.~Pagliara, and V.~Chandra, ``{A study of vorticity formation in high energy nuclear collisions},'' \href{http://dx.doi.org/10.1140/epjc/s10052-015-3624-1, 10.1140/epjc/s10052-018-5810-4}{{\em Eur. Phys. J.} {\bfseries C75} no.~9, (2015) 406}, \href{http://arxiv.org/abs/1501.04468}{{\ttfamily arXiv:1501.04468 [nucl-th]}}.
[Erratum: Eur. Phys. J.C78,no.5,354(2018)].

\bibitem{Becattini:2022zvf}
F.~Becattini, ``{Spin and polarization: a new direction in relativistic heavy ion physics},'' \href{http://dx.doi.org/10.1088/1361-6633/ac97a9}{{\em Rept. Prog. Phys.} {\bfseries 85} no.~12, (2022) 122301}, \href{http://arxiv.org/abs/2204.01144}{{\ttfamily arXiv:2204.01144 [nucl-th]}}.

\bibitem{DelZanna:2013eua}
L.~Del~Zanna, V.~Chandra, G.~Inghirami, V.~Rolando, A.~Beraudo, A.~De~Pace, G.~Pagliara, A.~Drago, and F.~Becattini, ``{Relativistic viscous hydrodynamics for heavy-ion collisions with ECHO-QGP},'' \href{http://dx.doi.org/10.1140/epjc/s10052-013-2524-5}{{\em Eur. Phys. J. C} {\bfseries 73} (2013) 2524}, \href{http://arxiv.org/abs/1305.7052}{{\ttfamily arXiv:1305.7052 [nucl-th]}}.

\bibitem{Karpenko:2013wva}
I.~Karpenko, P.~Huovinen, and M.~Bleicher, ``{A 3+1 dimensional viscous hydrodynamic code for relativistic heavy ion collisions},'' \href{http://dx.doi.org/10.1016/j.cpc.2014.07.010}{{\em Comput. Phys. Commun.} {\bfseries 185} (2014) 3016--3027}, \href{http://arxiv.org/abs/1312.4160}{{\ttfamily arXiv:1312.4160 [nucl-th]}}.

\bibitem{Ivanov:2019ern}
Y.~B. Ivanov, V.~D. Toneev, and A.~A. Soldatov, ``{Estimates of hyperon polarization in heavy-ion collisions at collision energies $\sqrt{s_{NN}}=$ 4--40 GeV},'' \href{http://dx.doi.org/10.1103/PhysRevC.100.014908}{{\em Phys. Rev. C} {\bfseries 100} no.~1, (2019) 014908}, \href{http://arxiv.org/abs/1903.05455}{{\ttfamily arXiv:1903.05455 [nucl-th]}}.

\bibitem{Li:2017slc}
H.~Li, L.-G. Pang, Q.~Wang, and X.-L. Xia, ``{Global $\Lambda$ polarization in heavy-ion collisions from a transport model},'' \href{http://dx.doi.org/10.1103/PhysRevC.96.054908}{{\em Phys. Rev.} {\bfseries C96} no.~5, (2017) 054908},
\href{http://arxiv.org/abs/1704.01507}{{\ttfamily arXiv:1704.01507 [nucl-th]}}.

\bibitem{Vitiuk:2019rfv}
O.~Vitiuk, L.~V. Bravina, and E.~E. Zabrodin, ``{Is different $\Lambda$ and $\bar \Lambda$ polarization caused by different spatio-temporal freeze-out picture?},'' \href{http://dx.doi.org/10.1016/j.physletb.2020.135298}{{\em Phys. Lett. B} {\bfseries 803} (2020) 135298}, \href{http://arxiv.org/abs/1910.06292}{{\ttfamily arXiv:1910.06292 [hep-ph]}}.

\bibitem{Sun:2017xhx}
Y.~Sun and C.~M. Ko, ``{$\Lambda$ hyperon polarization in relativistic heavy ion collisions from a chiral kinetic approach},'' \href{http://dx.doi.org/10.1103/PhysRevC.96.024906}{{\em Phys. Rev.} {\bfseries C96} no.~2, (2017) 024906},
\href{http://arxiv.org/abs/1706.09467}{{\ttfamily arXiv:1706.09467 [nucl-th]}}.

\bibitem{Becattini:2020ngo}
F.~Becattini and M.~A. Lisa, ``{Polarization and Vorticity in the Quark\textendash{}Gluon Plasma},'' \href{http://dx.doi.org/10.1146/annurev-nucl-021920-095245}{{\em Ann. Rev. Nucl. Part. Sci.} {\bfseries 70} (2020) 395--423}, \href{http://arxiv.org/abs/2003.03640}{{\ttfamily arXiv:2003.03640 [nucl-ex]}}.

\bibitem{Becattini:2021suc}
F.~Becattini, M.~Buzzegoli, and A.~Palermo, ``{Spin-thermal shear coupling in a relativistic fluid},'' \href{http://dx.doi.org/10.1016/j.physletb.2021.136519}{{\em Phys. Lett. B} {\bfseries 820} (2021) 136519}, \href{http://arxiv.org/abs/2103.10917}{{\ttfamily arXiv:2103.10917 [nucl-th]}}.

\bibitem{Liu:2021uhn}
S.~Y.~F. Liu and Y.~Yin, ``{Spin polarization induced by the hydrodynamic gradients},'' \href{http://dx.doi.org/10.1007/JHEP07(2021)188}{{\em JHEP} {\bfseries 07} (2021) 188}, \href{http://arxiv.org/abs/2103.09200}{{\ttfamily arXiv:2103.09200 [hep-ph]}}.

\bibitem{Becattini:2021iol}
F.~Becattini, M.~Buzzegoli, G.~Inghirami, I.~Karpenko, and A.~Palermo, ``{Local Polarization and Isothermal Local Equilibrium in Relativistic Heavy Ion Collisions},'' \href{http://dx.doi.org/10.1103/PhysRevLett.127.272302}{{\em Phys. Rev. Lett.} {\bfseries 127} no.~27, (2021) 272302}, \href{http://arxiv.org/abs/2103.14621}{{\ttfamily arXiv:2103.14621 [nucl-th]}}.

\bibitem{Fu:2021pok}
B.~Fu, S.~Y.~F. Liu, L.~Pang, H.~Song, and Y.~Yin, ``{Shear-Induced Spin Polarization in Heavy-Ion Collisions},'' \href{http://dx.doi.org/10.1103/PhysRevLett.127.142301}{{\em Phys. Rev. Lett.} {\bfseries 127} no.~14, (2021) 142301}, \href{http://arxiv.org/abs/2103.10403}{{\ttfamily arXiv:2103.10403 [hep-ph]}}.

\bibitem{Florkowski:2018fap}
W.~Florkowski, A.~Kumar, and R.~Ryblewski, ``{Relativistic hydrodynamics for spin-polarized fluids},'' \href{http://dx.doi.org/10.1016/j.ppnp.2019.07.001}{{\em Prog. Part. Nucl. Phys.} {\bfseries 108} (2019) 103709}, \href{http://arxiv.org/abs/1811.04409}{{\ttfamily arXiv:1811.04409 [nucl-th]}}.

\bibitem{Florkowski:2018ahw}
W.~Florkowski, A.~Kumar, and R.~Ryblewski, ``{Thermodynamic versus kinetic approach to polarization-vorticity coupling},'' \href{http://dx.doi.org/10.1103/PhysRevC.98.044906}{{\em Phys. Rev.} {\bfseries C98} (2018) 044906},
\href{http://arxiv.org/abs/1806.02616}{{\ttfamily arXiv:1806.02616 [hep-ph]}}.

\bibitem{Florkowski:2017dyn}
W.~Florkowski, B.~Friman, A.~Jaiswal, R.~Ryblewski, and E.~Speranza, ``{Spin-dependent distribution functions for relativistic hydrodynamics of spin-1/2 particles},'' \href{http://dx.doi.org/10.1103/PhysRevD.97.116017}{{\em Phys. Rev.} {\bfseries D97} no.~11, (2018) 116017},
\href{http://arxiv.org/abs/1712.07676}{{\ttfamily arXiv:1712.07676 [nucl-th]}}.

\bibitem{Florkowski:2017ruc}
W.~Florkowski, B.~Friman, A.~Jaiswal, and E.~Speranza, ``{Relativistic fluid dynamics with spin},'' \href{http://dx.doi.org/10.1103/PhysRevC.97.041901}{{\em Phys. Rev.} {\bfseries C97} no.~4, (2018) 041901},
\href{http://arxiv.org/abs/1705.00587}{{\ttfamily arXiv:1705.00587 [nucl-th]}}.

\bibitem{Florkowski:2019qdp}
W.~Florkowski, A.~Kumar, R.~Ryblewski, and R.~Singh, ``{Spin polarization evolution in a boost invariant hydrodynamical background},'' \href{http://dx.doi.org/10.1103/PhysRevC.99.044910}{{\em Phys. Rev.} {\bfseries C99} no.~4, (2019) 044910},
\href{http://arxiv.org/abs/1901.09655}{{\ttfamily arXiv:1901.09655 [hep-ph]}}.

\bibitem{Florkowski:2019voj}
W.~Florkowski, A.~Kumar, R.~Ryblewski, and A.~Mazeliauskas, ``{Longitudinal spin polarization in a thermal model},'' \href{http://dx.doi.org/10.1103/PhysRevC.100.054907}{{\em Phys. Rev.} {\bfseries C100} no.~5, (2019) 054907},
\href{http://arxiv.org/abs/1904.00002}{{\ttfamily arXiv:1904.00002 [nucl-th]}}.

\bibitem{Bhadury:2021oat}
S.~Bhadury, J.~Bhatt, A.~Jaiswal, and A.~Kumar, ``{New developments in relativistic fluid dynamics with spin},'' \href{http://dx.doi.org/10.1140/epjs/s11734-021-00020-4}{{\em Eur. Phys. J. ST} {\bfseries 230} no.~3, (2021) 655--672}, \href{http://arxiv.org/abs/2101.11964}{{\ttfamily arXiv:2101.11964 [hep-ph]}}.

\bibitem{Hattori:2019lfp}
K.~Hattori, M.~Hongo, X.-G. Huang, M.~Matsuo, and H.~Taya, ``{Fate of spin polarization in a relativistic fluid: An entropy-current analysis},'' \href{http://dx.doi.org/10.1016/j.physletb.2019.05.040}{{\em Phys. Lett.} {\bfseries B795} (2019) 100--106},
\href{http://arxiv.org/abs/1901.06615}{{\ttfamily arXiv:1901.06615 [hep-th]}}.

\bibitem{Fukushima:2020ucl}
K.~Fukushima and S.~Pu, ``{Spin hydrodynamics and symmetric energy-momentum tensors \textendash{} A current induced by the spin vorticity \textendash{}},'' \href{http://dx.doi.org/10.1016/j.physletb.2021.136346}{{\em Phys. Lett. B} {\bfseries 817} (2021) 136346}, \href{http://arxiv.org/abs/2010.01608}{{\ttfamily arXiv:2010.01608 [hep-th]}}.

\bibitem{Li:2020eon}
S.~Li, M.~A. Stephanov, and H.-U. Yee, ``{Nondissipative Second-Order Transport, Spin, and Pseudogauge Transformations in Hydrodynamics},'' \href{http://dx.doi.org/10.1103/PhysRevLett.127.082302}{{\em Phys. Rev. Lett.} {\bfseries 127} no.~8, (2021) 082302}, \href{http://arxiv.org/abs/2011.12318}{{\ttfamily arXiv:2011.12318 [hep-th]}}.

\bibitem{She:2021lhe}
D.~She, A.~Huang, D.~Hou, and J.~Liao, ``{Relativistic viscous hydrodynamics with angular momentum},'' \href{http://dx.doi.org/10.1016/j.scib.2022.10.020}{{\em Sci. Bull.} {\bfseries 67} (2022) 2265--2268}, \href{http://arxiv.org/abs/2105.04060}{{\ttfamily arXiv:2105.04060 [nucl-th]}}.

\bibitem{Montenegro:2017lvf}
D.~Montenegro, L.~Tinti, and G.~Torrieri, ``{Sound waves and vortices in a polarized relativistic fluid},'' \href{http://dx.doi.org/10.1103/PhysRevD.96.076016}{{\em Phys. Rev.} {\bfseries D96} no.~7, (2017) 076016},
\href{http://arxiv.org/abs/1703.03079}{{\ttfamily arXiv:1703.03079 [hep-th]}}.

\bibitem{Montenegro:2017rbu}
D.~Montenegro, L.~Tinti, and G.~Torrieri, ``{The ideal relativistic fluid limit for a medium with polarization},'' \href{http://dx.doi.org/10.1103/PhysRevD.96.056012}{{\em Phys. Rev.} {\bfseries D96} no.~5, (2017) 056012},
\href{http://arxiv.org/abs/1701.08263}{{\ttfamily arXiv:1701.08263 [hep-th]}}.

\bibitem{Florkowski:2018myy}
W.~Florkowski, E.~Speranza, and F.~Becattini, ``{Perfect-fluid hydrodynamics with constant acceleration along the stream lines and spin polarization},'' \href{http://dx.doi.org/10.5506/APhysPolB.49.1409}{{\em Acta Phys. Polon.} {\bfseries B49} (2018) 1409},
\href{http://arxiv.org/abs/1803.11098}{{\ttfamily arXiv:1803.11098 [nucl-th]}}.

\bibitem{Bhadury:2020puc}
S.~Bhadury, W.~Florkowski, A.~Jaiswal, A.~Kumar, and R.~Ryblewski, ``{Relativistic dissipative spin dynamics in the relaxation time approximation},'' \href{http://dx.doi.org/10.1016/j.physletb.2021.136096}{{\em Phys. Lett. B} {\bfseries 814} (2021) 136096}, \href{http://arxiv.org/abs/2002.03937}{{\ttfamily arXiv:2002.03937 [hep-ph]}}.

\bibitem{Shi:2020qrx}
S.~Shi, C.~Gale, and S.~Jeon, ``{From Chiral Kinetic Theory To Spin Hydrodynamics},'' \href{http://dx.doi.org/10.1016/j.nuclphysa.2020.121949}{{\em Nucl. Phys. A} {\bfseries 1005} (2021) 121949}, \href{http://arxiv.org/abs/2002.01911}{{\ttfamily arXiv:2002.01911 [nucl-th]}}.

\bibitem{Weickgenannt:2022zxs}
N.~Weickgenannt, D.~Wagner, E.~Speranza, and D.~H. Rischke, ``{Relativistic second-order dissipative spin hydrodynamics from the method of moments},'' \href{http://dx.doi.org/10.1103/PhysRevD.106.096014}{{\em Phys. Rev. D} {\bfseries 106} no.~9, (2022) 096014}, \href{http://arxiv.org/abs/2203.04766}{{\ttfamily arXiv:2203.04766 [nucl-th]}}.

\bibitem{Weickgenannt:2019dks}
N.~Weickgenannt, X.-L. Sheng, E.~Speranza, Q.~Wang, and D.~H. Rischke, ``{Kinetic theory for massive spin-1/2 particles from the Wigner-function formalism},''
\href{http://arxiv.org/abs/1902.06513}{{\ttfamily arXiv:1902.06513 [hep-ph]}}.

\bibitem{Weickgenannt:2020aaf}
N.~Weickgenannt, E.~Speranza, X.-l. Sheng, Q.~Wang, and D.~H. Rischke, ``{Generating Spin Polarization from Vorticity through Nonlocal Collisions},'' \href{http://dx.doi.org/10.1103/PhysRevLett.127.052301}{{\em Phys. Rev. Lett.} {\bfseries 127} no.~5, (2021) 052301}, \href{http://arxiv.org/abs/2005.01506}{{\ttfamily arXiv:2005.01506 [hep-ph]}}.

\bibitem{Singh:2024cub}
S.~K. Singh, R.~Ryblewski, and W.~Florkowski, ``{Spin dynamics with realistic hydrodynamic background for relativistic heavy-ion collisions},'' \href{http://dx.doi.org/10.1103/PhysRevC.111.024907}{{\em Phys. Rev. C} {\bfseries 111} no.~2, (2025) 024907}, \href{http://arxiv.org/abs/2411.08223}{{\ttfamily arXiv:2411.08223 [hep-ph]}}.

\bibitem{Sapna:2025yss}
Sapna, S.~K. Singh, and D.~Wagner, ``{Spin polarization of {\ensuremath{\Lambda}} hyperons from dissipative spin hydrodynamics},'' \href{http://dx.doi.org/10.1103/1s6g-fs8w}{{\em Phys. Rev. C} {\bfseries 112} no.~5, (2025) 054902}, \href{http://arxiv.org/abs/2503.22552}{{\ttfamily arXiv:2503.22552 [hep-ph]}}.

\bibitem{Dey:2024cwo}
S.~Dey and A.~Das, ``{Kubo formula for spin hydrodynamics: Spin chemical potential as the leading order term in a gradient expansion},'' \href{http://dx.doi.org/10.1103/PhysRevD.111.074037}{{\em Phys. Rev. D} {\bfseries 111} no.~7, (2025) 074037}, \href{http://arxiv.org/abs/2410.04141}{{\ttfamily arXiv:2410.04141 [nucl-th]}}.

\bibitem{Wang:2021ngp}
D.-L. Wang, S.~Fang, and S.~Pu, ``{Analytic solutions of relativistic dissipative spin hydrodynamics with Bjorken expansion},'' \href{http://dx.doi.org/10.1103/PhysRevD.104.114043}{{\em Phys. Rev. D} {\bfseries 104} no.~11, (2021) 114043}, \href{http://arxiv.org/abs/2107.11726}{{\ttfamily arXiv:2107.11726 [nucl-th]}}.

\bibitem{Daher:2022xon}
A.~Daher, A.~Das, W.~Florkowski, and R.~Ryblewski, ``{Canonical and phenomenological formulations of spin hydrodynamics},'' \href{http://dx.doi.org/10.1103/PhysRevC.108.024902}{{\em Phys. Rev. C} {\bfseries 108} no.~2, (2023) 024902}, \href{http://arxiv.org/abs/2202.12609}{{\ttfamily arXiv:2202.12609 [nucl-th]}}.

\bibitem{Sarwar:2022yzs}
G.~Sarwar, M.~Hasanujjaman, J.~R. Bhatt, H.~Mishra, and J.-e. Alam, ``{Causality and stability of relativistic spin hydrodynamics},'' \href{http://dx.doi.org/10.1103/PhysRevD.107.054031}{{\em Phys. Rev. D} {\bfseries 107} no.~5, (2023) 054031}, \href{http://arxiv.org/abs/2209.08652}{{\ttfamily arXiv:2209.08652 [nucl-th]}}.

\bibitem{Daher:2022wzf}
A.~Daher, A.~Das, and R.~Ryblewski, ``{Stability studies of first-order spin-hydrodynamic frameworks},'' \href{http://dx.doi.org/10.1103/PhysRevD.107.054043}{{\em Phys. Rev. D} {\bfseries 107} no.~5, (2023) 054043}, \href{http://arxiv.org/abs/2209.10460}{{\ttfamily arXiv:2209.10460 [nucl-th]}}.

\bibitem{Biswas:2022bht}
R.~Biswas, A.~Daher, A.~Das, W.~Florkowski, and R.~Ryblewski, ``{Boost invariant spin hydrodynamics within the first order in derivative expansion},'' \href{http://dx.doi.org/10.1103/PhysRevD.107.094022}{{\em Phys. Rev. D} {\bfseries 107} no.~9, (2023) 094022}, \href{http://arxiv.org/abs/2211.02934}{{\ttfamily arXiv:2211.02934 [nucl-th]}}.

\bibitem{Drogosz:2024lkx}
Z.~Drogosz, W.~Florkowski, N.~{\L}ygan, and R.~Ryblewski, ``{Boost-invariant spin hydrodynamics with spin feedback effects},'' \href{http://dx.doi.org/10.1103/PhysRevC.111.024909}{{\em Phys. Rev. C} {\bfseries 111} no.~2, (2025) 024909}, \href{http://arxiv.org/abs/2411.06154}{{\ttfamily arXiv:2411.06154 [hep-ph]}}.

\bibitem{Wang:2024afv}
D.-L. Wang, L.~Yan, and S.~Pu, ``{Late-time asymptotic solutions, attractor, and focusing behavior of spin hydrodynamics},'' \href{http://arxiv.org/abs/2408.03781}{{\ttfamily arXiv:2408.03781 [hep-ph]}}.

\bibitem{Florkowski:2010zz}
W.~Florkowski, {\em {Phenomenology of Ultra-Relativistic Heavy-Ion Collisions}}.
\newblock
2010.
\newblock

\bibitem{Vogt:2007zz}
R.~Vogt, {\em {Ultrarelativistic heavy-ion collisions}}.
\newblock Elsevier, Amsterdam, 2007.

\bibitem{Alam:1996fd}
J.~Alam, B.~Sinha, and S.~Raha, ``{Electromagnetic probes of quark gluon plasma},'' \href{http://dx.doi.org/10.1016/0370-1573(95)00084-4}{{\em Phys. Rept.} {\bfseries 273} (1996) 243--362}.

\bibitem{Kovtun:2019hdm}
P.~Kovtun, ``{First-order relativistic hydrodynamics is stable},'' \href{http://dx.doi.org/10.1007/JHEP10(2019)034}{{\em JHEP} {\bfseries 10} (2019) 034}, \href{http://arxiv.org/abs/1907.08191}{{\ttfamily arXiv:1907.08191 [hep-th]}}.

\bibitem{Biswas:2023qsw}
R.~Biswas, A.~Daher, A.~Das, W.~Florkowski, and R.~Ryblewski, ``{Relativistic second-order spin hydrodynamics: An entropy-current analysis},'' \href{http://dx.doi.org/10.1103/PhysRevD.108.014024}{{\em Phys. Rev. D} {\bfseries 108} no.~1, (2023) 014024}, \href{http://arxiv.org/abs/2304.01009}{{\ttfamily arXiv:2304.01009 [nucl-th]}}.

\bibitem{Rindori:2020qqa}
D.~Rindori, {\em {Entropy current in relativistic quantum statistical mechanics}}.
\newblock PhD thesis, U. Florence (main), Florence U., 2020.
\newblock \href{http://arxiv.org/abs/2110.07662}{{\ttfamily arXiv:2110.07662 [hep-th]}}.

\bibitem{Chen:2018cts}
S.~Coleman, \href{http://dx.doi.org/10.1142/9371}{{\em {Lectures of Sidney Coleman on Quantum Field Theory}}}.
\newblock WSP, Hackensack, 12, 2018.

\bibitem{HEHL197655}
F.~W. Hehl, ``On the energy tensor of spinning massive matter in classical field theory and general relativity,'' \href{http://dx.doi.org/https://doi.org/10.1016/0034-4877(76)90016-1}{{\em Reports on Mathematical Physics} {\bfseries 9} no.~1, (1976) 55--82}.

\bibitem{Speranza:2020ilk}
E.~Speranza and N.~Weickgenannt, ``{Spin tensor and pseudo-gauges: from nuclear collisions to gravitational physics},'' \href{http://dx.doi.org/10.1140/epja/s10050-021-00455-2}{{\em Eur. Phys. J. A} {\bfseries 57} no.~5, (2021) 155}, \href{http://arxiv.org/abs/2007.00138}{{\ttfamily arXiv:2007.00138 [nucl-th]}}.

\bibitem{BELINFANTE1939887}
F.~Belinfante, ``On the spin angular momentum of mesons,'' \href{http://dx.doi.org/https://doi.org/10.1016/S0031-8914(39)90090-X}{{\em Physica} {\bfseries 6} no.~7, (1939) 887--898}.

\bibitem{BELINFANTE1940449}
F.~Belinfante, ``On the current and the density of the electric charge, the energy, the linear momentum and the angular momentum of arbitrary fields,'' \href{http://dx.doi.org/https://doi.org/10.1016/S0031-8914(40)90091-X}{{\em Physica} {\bfseries 7} no.~5, (1940) 449--474}.

\bibitem{Rosenfeld1940}
L.~Rosenfeld {\em Mem. Acad. Roy. Bel.} {\bfseries 18} no.~1, (1940) .

\bibitem{DeGroot:1980dk}
S.~R. De~Groot, W.~A. Van~Leeuwen, and C.~G. Van~Weert, {\em {Relativistic Kinetic Theory. Principles and Applications}}.
\newblock North-Holland Publishing Company, 1980.

\bibitem{HILGEVOORD19631}
J.~Hilgevoord and S.~Wouthuysen, ``On the spin angular momentum of the dirac particle,'' \href{http://dx.doi.org/https://urldefense.com/v3/__https://doi.org/10.1016/0029-5582(63)90246-3__;!!DZ3fjg!s2gp30KOi9mvt-mJ6pIF2tdsyQlUCllEy_i1q_Bep6M5sf2Pf42G28WzXhzXFDJh$}{{\em Nuclear Physics} {\bfseries 40} (1963) 1 -- 12}.

\bibitem{HILGEVOORD19651002}
J.~Hilgevoord and E.~{De Kerf}, ``The covariant definition of spin in relativistic quantum field theory,'' \href{http://dx.doi.org/https://doi.org/10.1016/0031-8914(65)90141-2}{{\em Physica} {\bfseries 31} no.~7, (1965) 1002--1016}.

\bibitem{Weyssenhoff:1947iua}
J.~Weyssenhoff and A.~Raabe, ``{Relativistic dynamics of spin-fluids and spin-particles},''
{\em Acta Phys. Polon.} {\bfseries 9} (1947) 7--18.

\bibitem{Bjorken:1982qr}
J.~D. Bjorken, ``{Highly Relativistic Nucleus-Nucleus Collisions: The Central Rapidity Region},''
\href{http://dx.doi.org/10.1103/PhysRevD.27.140}{{\em Phys. Rev.} {\bfseries D27} (1983) 140--151}.

\bibitem{Daher:2024ixz}
A.~Daher, W.~Florkowski, and R.~Ryblewski, ``{Stability constraint for spin equation of state},'' \href{http://dx.doi.org/10.1103/PhysRevD.110.034029}{{\em Phys. Rev. D} {\bfseries 110} no.~3, (2024) 034029}, \href{http://arxiv.org/abs/2401.07608}{{\ttfamily arXiv:2401.07608 [hep-ph]}}.

\bibitem{FReif:1965}
F.~Reif, {\em {Fundamentals of Statistical and Thermal Physics}}.
\newblock 1965.

\bibitem{Green:1954}
M.~S. Green, ``{Markoff Random Processes and the Statistical Mechanics of Time‐Dependent Phenomena. II. Irreversible Processes in Fluids},'' \href{http://dx.doi.org/10.1063/1.1740082}{{\em J. Chem. Phys.} {\bfseries 22} (1954) 398}.

\bibitem{Kubo:1957mj}
R.~Kubo, ``{Statistical mechanical theory of irreversible processes. 1. General theory and simple applications in magnetic and conduction problems},'' \href{http://dx.doi.org/10.1143/JPSJ.12.570}{{\em J. Phys. Soc. Jap.} {\bfseries 12} (1957) 570--586}.

\bibitem{Hu:2021lnx}
J.~Hu, ``{Kubo formulae for first-order spin hydrodynamics},'' \href{http://dx.doi.org/10.1103/PhysRevD.103.116015}{{\em Phys. Rev. D} {\bfseries 103} no.~11, (2021) 116015}, \href{http://arxiv.org/abs/2101.08440}{{\ttfamily arXiv:2101.08440 [hep-ph]}}.

\bibitem{She:2024rnx}
D.~She, Y.-W. Qiu, and D.~Hou, ``{Relativistic second-order spin hydrodynamics: A Kubo-type formulation for~the quark-gluon plasma},'' \href{http://dx.doi.org/10.1103/PhysRevD.111.036027}{{\em Phys. Rev. D} {\bfseries 111} no.~3, (2025) 036027}, \href{http://arxiv.org/abs/2410.15142}{{\ttfamily arXiv:2410.15142 [nucl-th]}}.

\bibitem{Daher:2025pfq}
A.~Daher, X.-L. Sheng, D.~Wagner, and F.~Becattini, ``{Dissipative currents and transport coefficients in relativistic spin hydrodynamics},'' \href{http://arxiv.org/abs/2503.03713}{{\ttfamily arXiv:2503.03713 [nucl-th]}}.

\bibitem{Romatschke:2007mq}
P.~Romatschke and U.~Romatschke, ``{Viscosity Information from Relativistic Nuclear Collisions: How Perfect is the Fluid Observed at RHIC?},'' \href{http://dx.doi.org/10.1103/PhysRevLett.99.172301}{{\em Phys. Rev. Lett.} {\bfseries 99} (2007) 172301}, \href{http://arxiv.org/abs/0706.1522}{{\ttfamily arXiv:0706.1522 [nucl-th]}}.

\bibitem{Baier:2006um}
R.~Baier, P.~Romatschke, and U.~A. Wiedemann, ``{Dissipative hydrodynamics and heavy ion collisions},'' \href{http://dx.doi.org/10.1103/PhysRevC.73.064903}{{\em Phys. Rev. C} {\bfseries 73} (2006) 064903}, \href{http://arxiv.org/abs/hep-ph/0602249}{{\ttfamily arXiv:hep-ph/0602249}}.

\bibitem{Ryblewski:2015hea}
R.~Ryblewski and M.~Strickland, ``{Dilepton production from the quark-gluon plasma using (3+1)-dimensional anisotropic dissipative hydrodynamics},'' \href{http://dx.doi.org/10.1103/PhysRevD.92.025026}{{\em Phys. Rev. D} {\bfseries 92} no.~2, (2015) 025026}, \href{http://arxiv.org/abs/1501.03418}{{\ttfamily arXiv:1501.03418 [nucl-th]}}.

\bibitem{Vujanovic:2013jpa}
G.~Vujanovic, C.~Young, B.~Schenke, R.~Rapp, S.~Jeon, and C.~Gale, ``{Dilepton emission in high-energy heavy-ion collisions with viscous hydrodynamics},'' \href{http://dx.doi.org/10.1103/PhysRevC.89.034904}{{\em Phys. Rev. C} {\bfseries 89} no.~3, (2014) 034904}, \href{http://arxiv.org/abs/1312.0676}{{\ttfamily arXiv:1312.0676 [nucl-th]}}.

\bibitem{Bhatt:2011kx}
J.~R. Bhatt, H.~Mishra, and V.~Sreekanth, ``{Cavitation and thermal dilepton production in QGP},'' \href{http://dx.doi.org/10.1016/j.nuclphysa.2011.11.012}{{\em Nucl. Phys. A} {\bfseries 875} (2012) 181--196}, \href{http://arxiv.org/abs/1101.5597}{{\ttfamily arXiv:1101.5597 [hep-ph]}}.

\bibitem{Singh:2018bih}
B.~Singh, J.~R. Bhatt, and H.~Mishra, ``{Probing vorticity in heavy ion collision with dilepton production},'' \href{http://dx.doi.org/10.1103/PhysRevD.100.014016}{{\em Phys. Rev. D} {\bfseries 100} no.~1, (2019) 014016}, \href{http://arxiv.org/abs/1811.08124}{{\ttfamily arXiv:1811.08124 [hep-ph]}}.

\bibitem{Bailhache:2025kwa}
R.~Bailhache and H.~Appelsh{\"a}user, ``{Dileptons at Colliders as Probes of the Quark{\textendash}Gluon Plasma},'' \href{http://dx.doi.org/10.1146/annurev-nucl-121423-100858}{{\em Ann. Rev. Nucl. Part. Sci.} {\bfseries 75} no.~1, (2025) 463--486}, \href{http://arxiv.org/abs/2512.10597}{{\ttfamily arXiv:2512.10597 [nucl-ex]}}.

\bibitem{Meninno:2020mjd}
{\bfseries ALICE} Collaboration, E.~Meninno, ``{Dilepton measurements with the ALICE experiment at the LHC},'' \href{http://dx.doi.org/10.1088/1742-6596/1602/1/012016}{{\em J. Phys. Conf. Ser.} {\bfseries 1602} no.~1, (2020) 012016}.

\bibitem{STAR:2015tnn}
{\bfseries STAR} Collaboration, L.~Adamczyk {\em et~al.}, ``{Measurements of Dielectron Production in Au$+$Au Collisions at $\sqrt{s_{\rm NN}}$ = 200 GeV from the STAR Experiment},'' \href{http://dx.doi.org/10.1103/PhysRevC.92.024912}{{\em Phys. Rev. C} {\bfseries 92} no.~2, (2015) 024912}, \href{http://arxiv.org/abs/1504.01317}{{\ttfamily arXiv:1504.01317 [hep-ex]}}.

\bibitem{Song:2018xca}
T.~Song, W.~Cassing, P.~Moreau, and E.~Bratkovskaya, ``{Open charm and dileptons from relativistic heavy-ion collisions},'' \href{http://dx.doi.org/10.1103/PhysRevC.97.064907}{{\em Phys. Rev. C} {\bfseries 97} no.~6, (2018) 064907}, \href{http://arxiv.org/abs/1803.02698}{{\ttfamily arXiv:1803.02698 [nucl-th]}}.

\bibitem{Song:2018dvf}
T.~Song, W.~Cassing, P.~Moreau, and E.~Bratkovskaya, ``{Discrepancy in low transverse momentum dileptons from relativistic heavy-ion collisions},'' \href{http://dx.doi.org/10.1103/PhysRevC.98.041901}{{\em Phys. Rev. C} {\bfseries 98} no.~4, (2018) 041901}, \href{http://arxiv.org/abs/1806.09377}{{\ttfamily arXiv:1806.09377 [nucl-th]}}.

\bibitem{Song:2018wvd}
T.~Song and P.~Moreau, ``{Low-energy Bremsstrahlung photon in relativistic nucleon+nucleon collisions},'' \href{http://dx.doi.org/10.1103/PhysRevD.98.116007}{{\em Phys. Rev. D} {\bfseries 98} no.~11, (2018) 116007}, \href{http://arxiv.org/abs/1810.08013}{{\ttfamily arXiv:1810.08013 [nucl-th]}}.

\bibitem{Rapp:2013nxa}
R.~Rapp, ``{Dilepton Spectroscopy of QCD Matter at Collider Energies},'' \href{http://dx.doi.org/10.1155/2013/148253}{{\em Adv. High Energy Phys.} {\bfseries 2013} (2013) 148253}, \href{http://arxiv.org/abs/1304.2309}{{\ttfamily arXiv:1304.2309 [hep-ph]}}.

\bibitem{Rapp:1999ej}
R.~Rapp and J.~Wambach, ``{Chiral symmetry restoration and dileptons in relativistic heavy ion collisions},'' \href{http://dx.doi.org/10.1007/0-306-47101-9_1}{{\em Adv. Nucl. Phys.} {\bfseries 25} (2000) 1}, \href{http://arxiv.org/abs/hep-ph/9909229}{{\ttfamily arXiv:hep-ph/9909229}}.

\bibitem{Wang:2018sur}
X.~Wang, M.~Wei, Z.~Li, and M.~Huang, ``{Quark matter under rotation in the NJL model with vector interaction},'' \href{http://dx.doi.org/10.1103/PhysRevD.99.016018}{{\em Phys. Rev. D} {\bfseries 99} no.~1, (2019) 016018}, \href{http://arxiv.org/abs/1808.01931}{{\ttfamily arXiv:1808.01931 [hep-ph]}}.

\bibitem{Zhu:2025pxh}
Z.-B. Zhu, H.-L. Chen, and X.-G. Huang, ``{Chiral symmetry breaking in accelerating and rotating frames},'' \href{http://dx.doi.org/10.1103/yxj9-33z6}{{\em Phys. Rev. D} {\bfseries 113} no.~3, (2026) 034005}, \href{http://arxiv.org/abs/2511.03230}{{\ttfamily arXiv:2511.03230 [hep-ph]}}.

\bibitem{Farias:2025vss}
R.~L.~S. Farias and W.~R. Tavares, ``{Chiral and deconfinement transitions in spin-polarized quark matter},'' \href{http://dx.doi.org/10.1103/5znc-7ztg}{{\em Phys. Rev. D} {\bfseries 112} no.~5, (2025) L051902}, \href{http://arxiv.org/abs/2507.08130}{{\ttfamily arXiv:2507.08130 [hep-ph]}}.

\bibitem{Pradhan:2025pol}
K.~K. Pradhan, D.~Sahu, and R.~Sahoo, ``{Emergent spin polarization from {\ensuremath{\rho}} meson condensation in rotating hadronic matter},'' \href{http://dx.doi.org/10.1016/j.physletb.2025.140090}{{\em Phys. Lett. B} {\bfseries 872} (2026) 140090}, \href{http://arxiv.org/abs/2510.22755}{{\ttfamily arXiv:2510.22755 [hep-ph]}}.

\bibitem{Mukherjee:2023ijv}
G.~Mukherjee, D.~Dutta, and D.~K. Mishra, ``{Conserved number fluctuations under global rotation in a hadron resonance gas model},'' \href{http://dx.doi.org/10.1140/epjc/s10052-024-12592-1}{{\em Eur. Phys. J. C} {\bfseries 84} no.~3, (2024) 258}, \href{http://arxiv.org/abs/2304.14658}{{\ttfamily arXiv:2304.14658 [hep-ph]}}.

\bibitem{Wu:2019eyi}
H.-Z. Wu, L.-G. Pang, X.-G. Huang, and Q.~Wang, ``{Local spin polarization in high energy heavy ion collisions},'' \href{http://dx.doi.org/10.1103/PhysRevResearch.1.033058}{{\em Phys. Rev. Research.} {\bfseries 1} (2019) 033058},
\href{http://arxiv.org/abs/1906.09385}{{\ttfamily arXiv:1906.09385 [nucl-th]}}.

\end{thebibliography}\endgroup
\bibliographystyle{utphys}
\end{document}